\documentclass{article}
\usepackage{iclr2025_conference,times} 

\usepackage{amsmath,amsfonts,bm}

\def\eqref#1{equation~\ref{#1}}

\def\1{\bm{1}}

\DeclareMathAlphabet{\mathsfit}{\encodingdefault}{\sfdefault}{m}{sl}
\SetMathAlphabet{\mathsfit}{bold}{\encodingdefault}{\sfdefault}{bx}{n}

\usepackage[T1]{fontenc}

\usepackage{hyperref}
\usepackage{url}
\usepackage{multirow}
\usepackage{graphicx}
\usepackage{wrapfig}
\usepackage{amssymb}
\usepackage{lipsum}

\usepackage{amsfonts}

\usepackage{pifont}

\usepackage{wrapfig}
\usepackage{makecell}

\usepackage{float}

\def\our{GS-Verse}

\title{\our{}: Mesh-based Gaussian Splatting for Physics-aware Interaction in Virtual Reality}

\author{Anastasiya Pechko, Piotr Borycki, Joanna Waczy\'nska, Daniel Barczyk, Agata Szyma\'nska \\
Jagiellonian University\\
\texttt{anastasiya.pechko@doctoral.uj.edu.pl} \\
\And
S{\l}awomir Tadeja \\
University of Cambridge \\
\And
Przemys{\l}aw Spurek \\
Jagiellonian University \\
IDEAS Research Institute \\
\texttt{przemyslaw.spurek@uj.edu.com} \\
}

\iclrfinalcopy 
\begin{document}

\maketitle

 \begin{figure}[h!]
     \centering
       \vspace{-0.4cm} \includegraphics[width=1\textwidth]{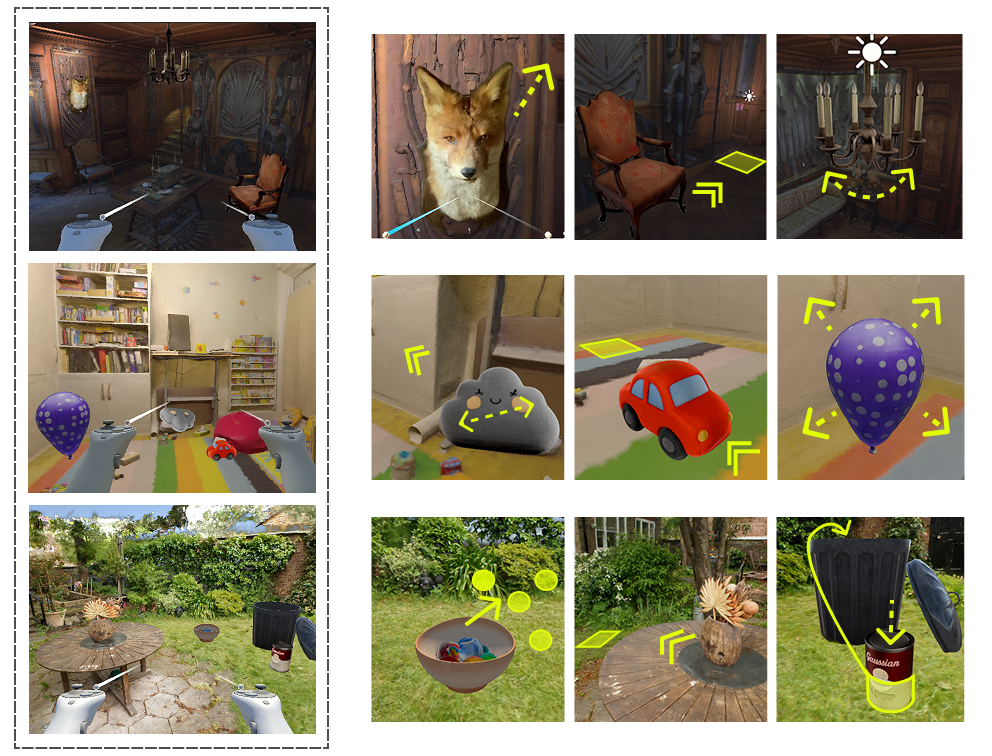}
      \vspace{-0.6cm} 
     \caption{The three scenes used in our study: (top) \textit{dark room}, (middle) \textit{toy room}, and (bottom) \textit{garden} as seen from VR. The three columns on the right present a handful from a large range of possible physics-aware 3D object manipulations (e.g., moving, swinging, stretching, pulling, twisting, shaking, crushing, tipping).
     } 
\label{fig:teaser}
 \end{figure}

\begin{abstract}
As the demand for immersive 3D content grows, the need for intuitive and efficient interaction methods becomes paramount. Current techniques for physically manipulating 3D content within Virtual Reality (VR) often face significant limitations, including reliance on engineering-intensive processes and simplified geometric representations, such as tetrahedral cages, which can compromise visual fidelity and physical accuracy. In this paper, we introduce \our{} (\textbf{G}aussian \textbf{S}platting for \textbf{V}irtual \textbf{E}nvironment \textbf{R}endering and \textbf{S}cene \textbf{E}diting), a novel method designed to overcome these challenges by directly integrating an object's mesh with a Gaussian Splatting (GS) representation. Our approach enables more precise surface approximation, leading to highly realistic deformations and interactions. By leveraging existing 3D mesh assets, \our{} facilitates seamless content reuse and simplifies the development workflow.
Moreover, our system is designed to be physics-engine-agnostic, granting developers robust deployment flexibility. This versatile architecture delivers a highly realistic, adaptable, and intuitive approach to interactive 3D manipulation. We rigorously validate our method against the current state-of-the-art technique that couples VR with GS in a comparative user study involving 18 participants. Specifically, we demonstrate that our approach is statistically significantly better for physics-aware stretching manipulation and is also more consistent in other physics-based manipulations like twisting and shaking. Further evaluation across various interactions and scenes confirms that our method consistently delivers high and reliable performance, showing its potential as a plausible alternative to existing methods. Code is available at \url{https://github.com/Anastasiya999/GS-Verse}
\end{abstract}

\section{Introduction}


The creation of immersive 3D environments has traditionally been the domain of skilled experts and graphic artists~\cite{huang_personalized_2025}. Conventional workflows, reliant on meticulous mesh modeling and texture mapping, are labor-intensive, time-consuming, and require a high level of technical expertise \cite{yuan2025immersegen, barron2022mip}. This complexity poses a significant barrier to the widespread adoption and creation of rich, interactive 3D content, particularly for emerging platforms like \textit{Virtual Reality} (VR) \cite{tadeja_2020, huang_personalized_2025}.

An alternative paradigm has recently emerged with \textit{3D Gaussian Splatting} (3DGS) \cite{kerbl3Dgaussians}, a technique that enables the creation of photorealistic 3D assets and entire scenes, which can be used to populate immersive environments \cite{qiu2025advancing}. This approach enables the reconstruction of detailed 3D objects and scenes from video recordings, typically captured with a standard smartphone. This breakthrough allows experts and non-experts alike to develop VR environments or capture and import individual real-world objects into existing digital scenes, significantly lowering the barrier to entry for content creators \cite{qiu2025advancing}.

GS has already been adopted for VR applications, with recent work demonstrating its integration into popular development platforms such as Unity game engine~\cite{jiang2024vr,franke2025vr,tu2025vrsplat}. VR-Splatting introduces a foveated rendering framework that combines splats with neural points to reduce computational cost by adapting rendering quality to the user’s gaze while preserving fidelity in the foveal region~\cite{franke2025vr}. Similarly, VRSplat presents a fast and robust pipeline tailored for VR, emphasizing low-latency and stable performance to enable interactive experiences~\cite{tu2025vrsplat}. VR-GS incorporates physics-aware dynamics into GS, allowing interactive manipulation of splatted objects in immersive environments~\cite{jiang2024vr}. However, this approach has a notable limitation--it relies on simplified geometric proxies, such as low-resolution tetrahedral cages, to approximate the object's physics. This simplification, although computationally efficient, compromises the accuracy of physical simulations and can result in visually unconvincing deformations that do not accurately represent the object's true surface geometry.

In this paper, we introduce \our{} which addresses the key challenges of crafting, editing, and interacting with digital assets by leveraging the strengths of GS within an immersive environment, see Fig.~\ref{fig:teaser}. While previous work has made strides in integrating physical simulations with GS, they often rely on simplified geometric representations (e.g., tetrahedral cages), which can compromise the visual fidelity of the final rendering and the plausibility of physical interactions.

\begin{figure}[t]
  \centering
  \includegraphics[width=1.0\textwidth]{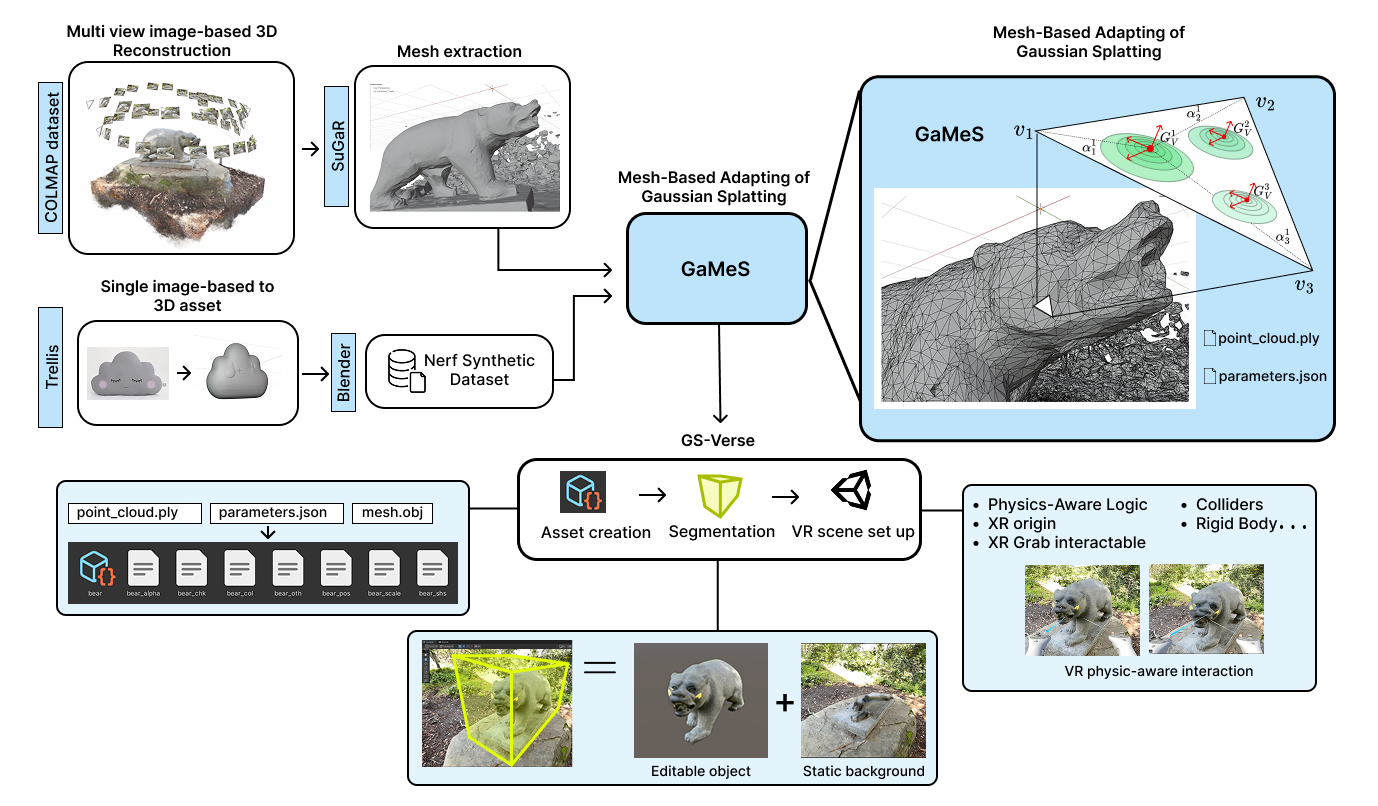}
  \caption{Overview of our proposed method, \our{}. 
The approach enables real-time interaction in VR by generating mesh-based Gaussian Splatting assets. 
The processing pipeline begins with multiview image scene reconstruction, followed by mesh extraction using \textit{SuGaR} or \textit{Trellis}. 
A subsequent segmentation step optimizes the scene by dividing it into dynamic and static components. 
Thanks to mesh-based parameterization via \textit{GaMeS}, the resulting representations can be seamlessly integrated into physics-aware engines such as Unity, enabling efficient and physically consistent VR interactions.}
  \label{fig:method}
\end{figure}

\begin{figure}[t]
    \centering
    \begin{tabular}{c c c c c}
        Method & stretch & twist & shake & static\\
        \raisebox{2\normalbaselineskip}[0pt][0pt]{\rotatebox{90}{VR-GS}} & 
        \includegraphics[width=0.2\textwidth]{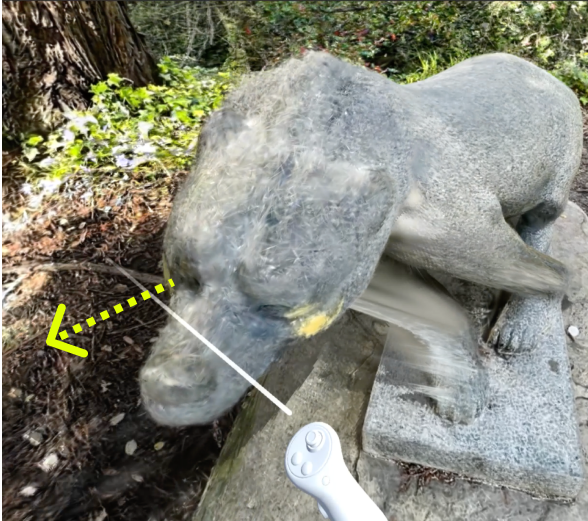} & 
        \includegraphics[width=0.2\textwidth]{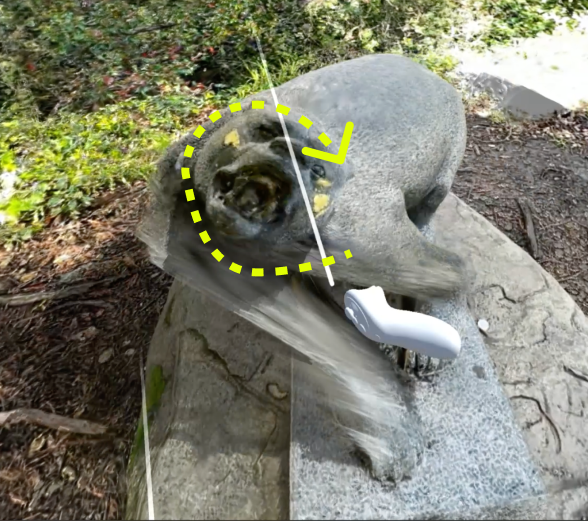} & 
        \includegraphics[width=0.2\textwidth]{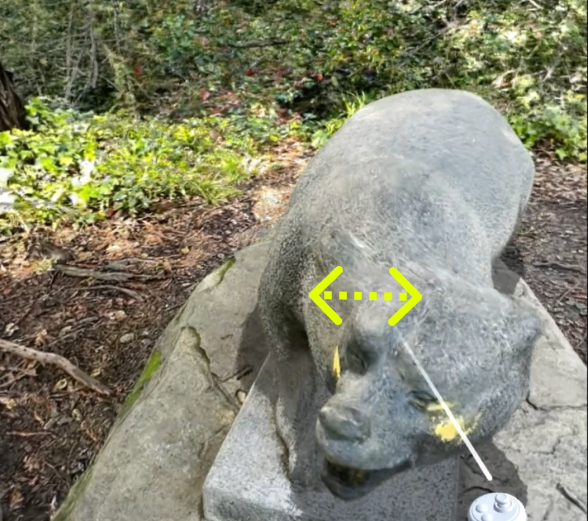} & 
        \includegraphics[width=0.2\textwidth]{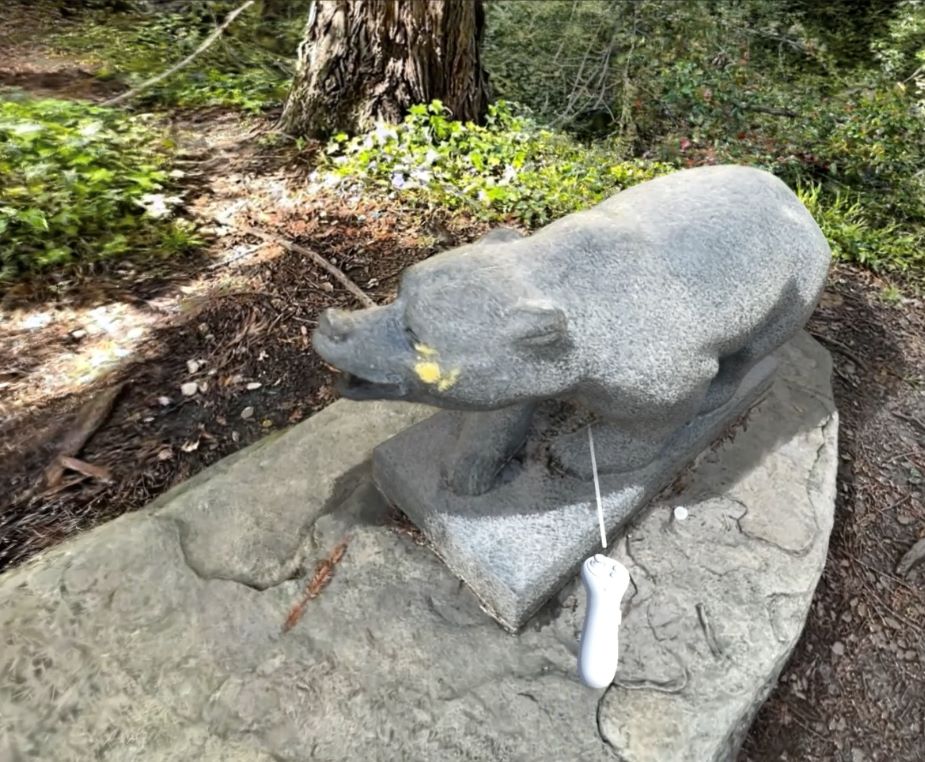} \\
        
        \raisebox{2\normalbaselineskip}[0pt][0pt]{\rotatebox{90}{\our{}}} & 
        \includegraphics[width=0.2\textwidth]{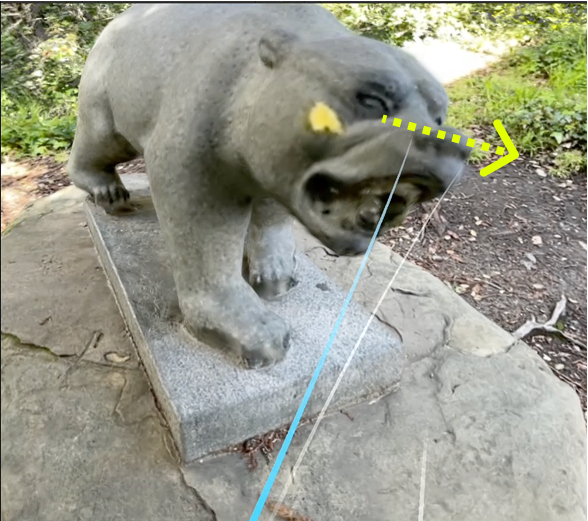} & 
        \includegraphics[width=0.2\textwidth]{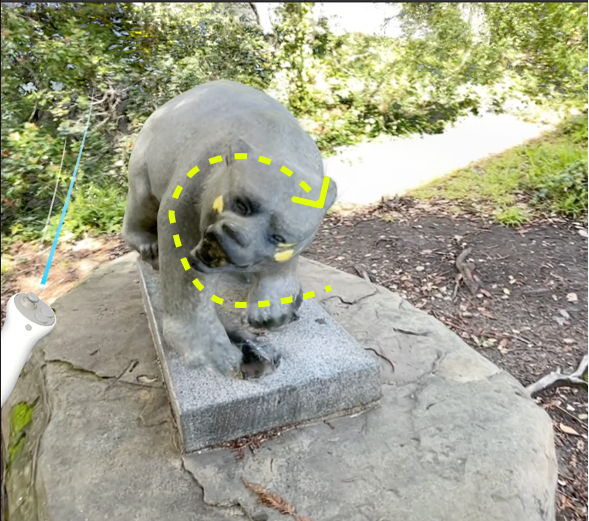} & 
        \includegraphics[width=0.2\textwidth]{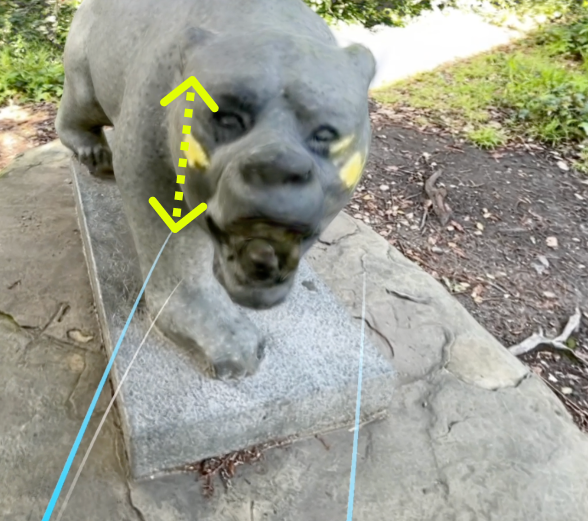} & 
        \includegraphics[width=0.2\textwidth]{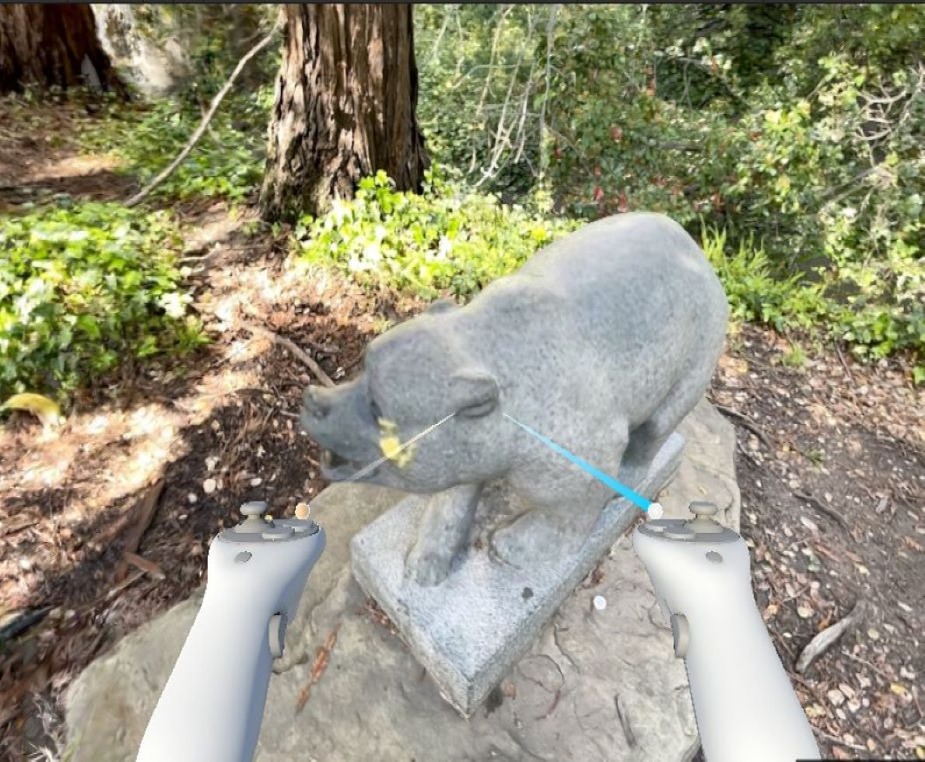}
    \end{tabular}
    \caption{Visual example of a participant performing the requested manipulations during the first, closed-ended task: stretch, twist, shake in  VR-GS (first row) and \our{} method (second row).}
    \label{fig:vrgs_comp_first_scene}
\end{figure}

\begin{figure}[t]
    \centering
    \begin{tabular}{c c c c c}
         &  \multicolumn{4}{c}{Example modification}\\
        \raisebox{2\normalbaselineskip}[0pt][0pt]{\rotatebox{90}{VR-GS}} & 
        \includegraphics[width=0.2\textwidth]{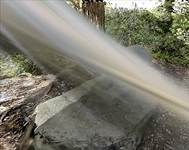} & 
        \includegraphics[width=0.2\textwidth]{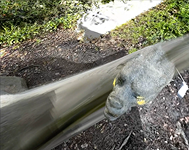} & 
        \includegraphics[width=0.2\textwidth]{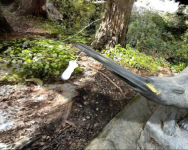} & 
        \includegraphics[width=0.2\textwidth]{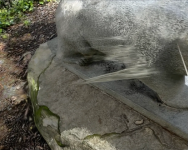} \\
        
        \raisebox{2\normalbaselineskip}[0pt][0pt]{\rotatebox{90}{\our{}}} & 
        \includegraphics[width=0.2\textwidth]{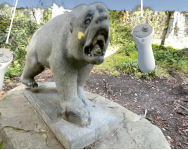} & 
        \includegraphics[width=0.2\textwidth]{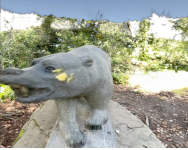} & 
        \includegraphics[width=0.2\textwidth]{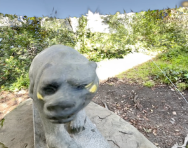} & 
        \includegraphics[width=0.2\textwidth]{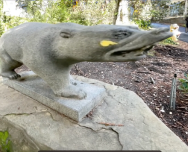}
    \end{tabular}
    \caption{ Artifacts during maximum object stretching from the user perspective. In VR-GS (top row), overstretching caused very large splats that overlapped the object. In addition, users were able to detach individual splats, making the interaction unsuccessful. In contrast, our \our{} (bottom row) remained robust even under extreme stretching, showing minimal artifacts.}
    \label{fig:beararteffact}
\end{figure}

Our model represents a novel approach that directly employs the object's surface mesh as the primary representation for physical simulation, see Fig.~\ref{fig:method}. In practice, we can estimate high-quality meshes directly from GS representation \cite{guedon2024sugar, waczynska2024games, huang20242d, li2025geosvr}, which allows us to model large 3D scenes. For content generation, we can utilize similar tools to convert 2D images into GS representations and extract meshes. Additionally, objects can also be obtained using specialized generative models, which create meshes with GS representations \cite{xiang2025structured}. 
The use of high-quality surface meshes offers several key advantages. First, it ensures a more accurate approximation of the object’s geometry, resulting in more realistic deformations and interactions, see Fig.~\ref{fig:vrgs_comp_first_scene}. Second, by utilizing a standard mesh, our system is compatible with existing 3D models and assets, allowing for the seamless reuse of a vast array of interactive VR applications, see Fig.~\ref{fig:beararteffact}. 

Furthermore, our system is designed to be engine-agnostic, providing a flexible framework that can be integrated with any off-the-shelf physics engine. This versatility allows developers to choose the simulation tool best suited for their specific needs, without being locked into a proprietary or specialized solution. The ability to use a wide range of physics engines not only simplifies the development workflow but also broadens the potential applications of our system. By combining these unique advantages, our \our{} method offers a robust, flexible, and highly realistic solution for immersive 3D content manipulation.

The following constitutes a list of our key contributions:

\begin{enumerate}
    \item We propose \our{} pipeline that uses surface meshes with GS representation in VR, ensuring more accurate object geometry approximation and backward compatibility with the majority of existing game engine systems.
    \item \our{} is engine-agnostic and works with existing physics simulation engines to provide optimized, physics-aware interaction and manipulation, contributing to more natural and immersive VR experiences.
    \item \our{} offers highly realistic, adaptable 3D manipulation, providing a statistically significant advancement in physics-aware stretching and consistent perceived performance across diverse manipulation scenarios.
    
\end{enumerate}

\section{Related works}

Here, we review the key areas of research that provide the foundation for our work. We structure our discussion into two main parts. First, we provide an overview of radiance fields, focusing on the evolution from Neural Radiance Fields (NeRF) to modern point-based representations like 3D GS. Second, we examine the specific challenges and existing solutions for using Radiance Fields in Virtual Reality.

\paragraph{Radiance Fields} 

The challenge of synthesizing novel views from a collection of images has long been a central focus in computer graphics. A significant breakthrough in this area occurred with the introduction of NeRFs \cite{mildenhall2021nerf}, which combine the principles of classical volume rendering with the expressive power of multilayer perceptrons (MLPs) to produce images of remarkable quality. Despite their visual fidelity, NeRFs are inherently limited by substantial computational costs, as their reliance on per-pixel ray marching results in prolonged training and rendering times. In response, a significant body of research has emerged to address these performance bottlenecks \cite{fridovich2022plenoxels,muller2022instant,steiner2024frustum,zimny2025multiplanenerf}, extend the framework to large, unbounded scenes \cite{barron2022mip}, and adapt it for manual editing and physical engines \cite{chen2023neuraleditor, wang2023rip, zielinski2025genie}.

To overcome the NeRFs' performance limitations, 3D Gaussian Splatting (3DGS) \cite{kerbl3Dgaussians} introduced a paradigm shift by representing scenes with an explicit set of 3D Gaussians. This approach leverages highly efficient software rasterization, resulting in a significant acceleration in rendering speed. The success of 3DGS has spurred a new wave of research, with efforts focused on enhancing the technique's robustness and efficiency. This includes developing methods for artifact-free rendering \cite{huang2024error,radl2024stopthepop,yu2024mip}, strategies for reducing the number of primitives required to represent a scene without compromising quality \cite{fan2024lightgaussian,fang2024mini}, and scaling the representation to handle massive datasets for large-scale exploration \cite{lin2024vastgaussian}. GS representations are well-suited for manual editing and physical engines \cite{waczynska2024games,guedon2024sugar,borycki2024gasp,tobiasz2025meshsplats}

\paragraph{Radiance Fields for VR}

Immersive interfaces such as VR serve as a natural and compelling application for radiance fields, offering users an unparalleled sense of immersion when exploring captured scenes. Early efforts to integrate NeRFs into VR, such as FoV-NeRF \cite{deng2022fov}, utilized gaze-contingent neural representations and foveated rendering to achieve real-time performance. Another approach, VR-NeRF \cite{xu2023vr}, focuses on architectural solutions, distributing the computational load across multiple GPUs and utilizing occupancy grids to accelerate rendering. However, these methods still struggle to meet the demanding performance targets of VR without significant compromises in visual quality.

The advent of 3DGS has opened new avenues for VR applications. For instance, VR-GS \cite{jiang2024vr} demonstrated how an interactive layer can be built upon the 3DGS framework to enable physics-based manipulation of objects within the scene. While innovative, its reliance on simplified proxy geometries for simulation is a limitation that our proposed method directly addresses, positioning our work as a potential high-quality replacement. Other concurrent research has explored foveated rendering for GS on mobile hardware  \cite{lin2024rtgs}, proposing a level-of-detail strategy and a training regimen designed to reduce the overlap between Gaussians. More recently, VRSplat \cite{tu2025vrsplat} introduced a fast and robust pipeline specifically designed for VR, featuring a novel anti-aliasing method to mitigate visual artifacts like popping and flickering. VR-Splatting \cite{franke2025vr} proposed a hybrid foveated system that elegantly combines the strengths of neural point rendering for the sharp foveal region with the smoothness of 3DGS for the periphery, successfully meeting VR's performance demands. While these works significantly advance rendering efficiency, our research diverges by focusing on the core challenge of high-fidelity physical interaction in VR Gaussian Splatting, which remains underexplored and is rarely supported by publicly available implementations.

\section{Mesh-based GS for Physics-aware Interaction in VR (\our{})}

\paragraph{Gaussian Splatting} 
GS represents a 3D scene using a collection of spatially distributed 3D Gaussian primitives. Each Gaussian is defined by its mean position, covariance matrix, opacity value, and color encoded through spherical harmonics (SH)~\cite{fridovich2022plenoxels, muller2022instant}.

The GS approach constructs a radiance field by iteratively optimizing the positions, covariances, opacities, and SH-based color coefficients of the Gaussians. One of the key advantages of GS lies in its rendering pipeline, which efficiently projects 3D Gaussian components onto 2D image planes, resulting in a highly efficient rendering pipeline.

The 3D scene is modeled as a dense set of Gaussians:
\[
\mathcal{G} = \left\{ \left( \mathcal{N}(\mathbf{m}_i, \Sigma_i), \sigma_i, c_i \right) \right\}_{i=1}^{n},
\]

where $\mathbf{m}_i$ is the mean (position), $\Sigma_i$ is the covariance matrix, $\sigma_i$ represents the opacity, and $c_i$ denotes the SH color of the $i$-th Gaussian.

Training involves rendering from the current set of Gaussians and comparing the synthesized views to ground-truth images. Because projecting from 3D to 2D may introduce geometric errors, the optimization is capable of adding, removing, or relocating Gaussians to correct such inaccuracies.

To adaptively refine the scene representation, a \textit{densification strategy} is employed during training. New Gaussians are created by splitting existing ones based on heuristics such as gradient magnitude, visibility, and accumulated opacity. This allows the model to allocate more representational capacity in regions with high complexity, while avoiding unnecessary overhead in simpler areas.

The optimization process minimizes a \textit{photometric reconstruction loss}, typically defined as the mean squared error (MSE) between rendered images and the ground-truth views. This is often combined with additional regularization terms that encourage spatial compactness, control Gaussian growth, and stabilize the covariance matrices during training.

\begin{figure}[t]
    \centering
    \begin{tabular}{c c c c}
        Method & Simulation Mesh & Original Render & VR View \\
        \raisebox{2\normalbaselineskip}[0pt][0pt]{\rotatebox{90}{VR-GS}} & 
        \includegraphics[width=0.25\textwidth]{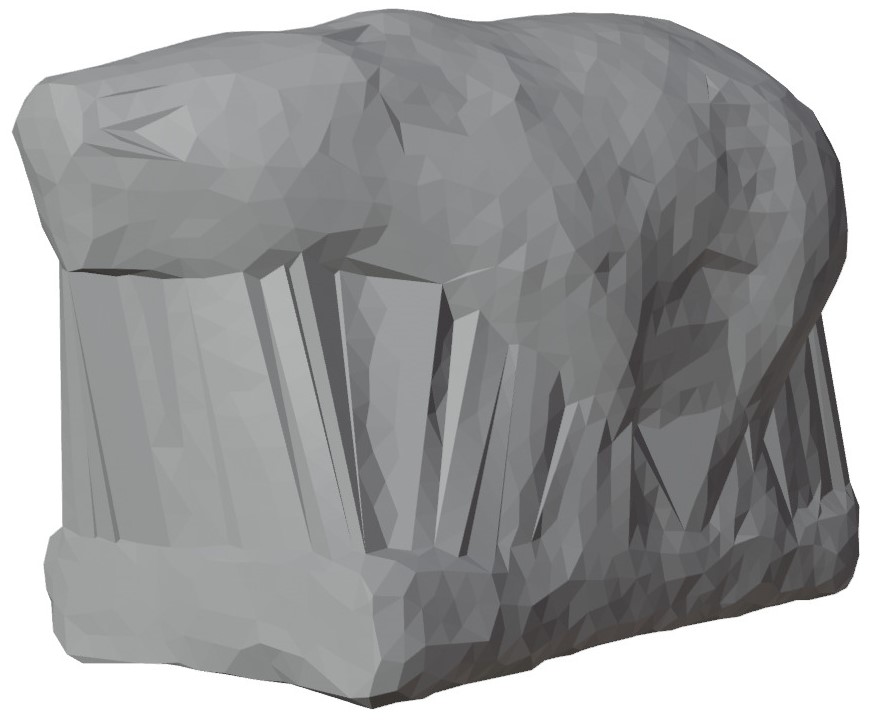} & 
        \includegraphics[width=0.25\textwidth]{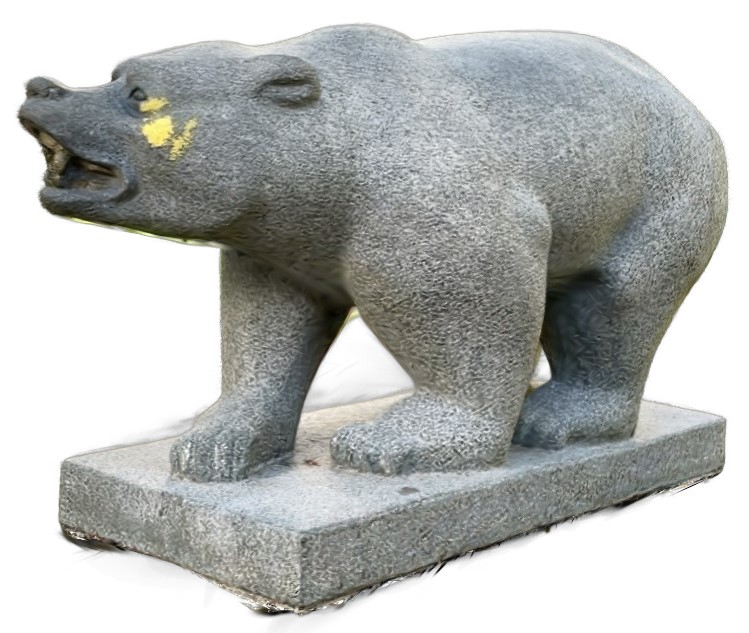} & 
        \includegraphics[width=0.25\textwidth]{imgs/vrgs_comp/vrgs_vr_orig.jpg} \\
        
        \raisebox{2\normalbaselineskip}[0pt][0pt]{\rotatebox{90}{\our{}}} & 
        \includegraphics[width=0.25\textwidth]{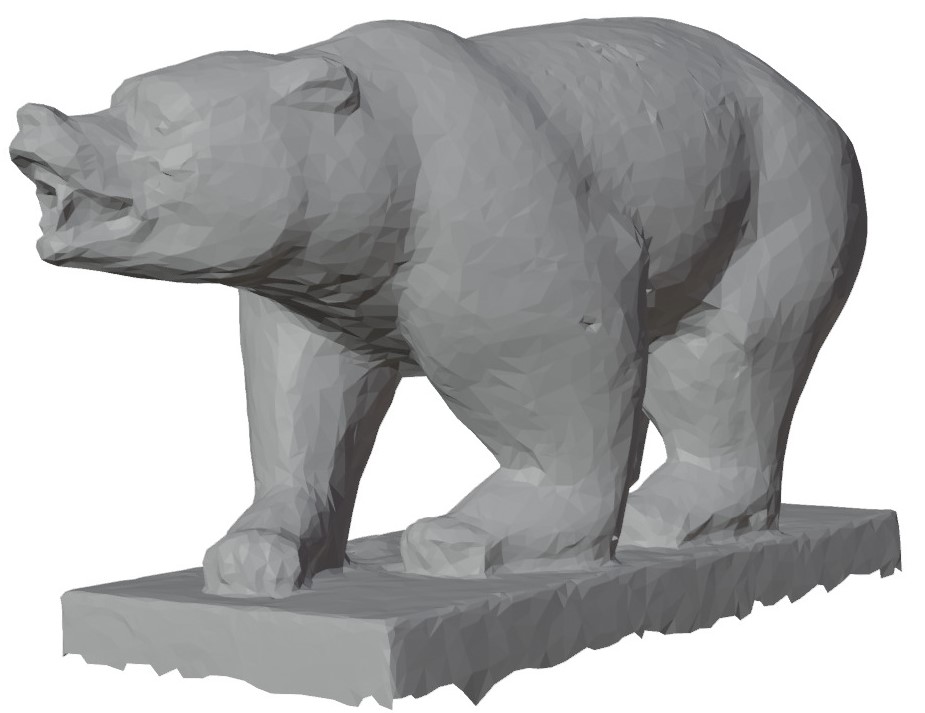} & 
        \includegraphics[width=0.25\textwidth]{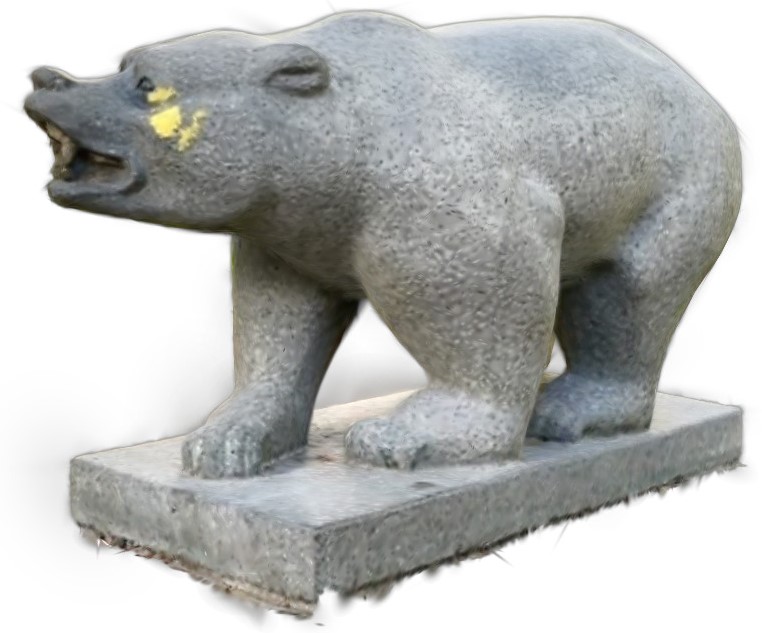} & 
        \includegraphics[width=0.25\textwidth]{imgs/vrgs_comp/our_vr_orig.jpg}
    \end{tabular}
    \caption{Comparison of VR-GS and \our{} across three visualization stages: the simulation mesh (left), the original render (middle), and the render inside VR (right). Unlike VR-GS, which relies on a cage mesh, \our{} employs a more geometry-accurate simulation mesh.}
    \label{fig:vrgs_comparison}
\end{figure}

\paragraph{Mesh Reconstruction for GS}  
An essential step in extending GS beyond photorealistic rendering is the recovery of a consistent and geometrically accurate mesh representation. Recent methods demonstrate that imposing surface constraints and geometric regularization significantly improve the fidelity of reconstructed meshes. For instance, SuGaR \cite{guedon2024sugar} introduces surface-aligned splats that facilitate efficient and accurate mesh extraction, while 2DGS \cite{huang20242d} leverages 2D Gaussians embedded in the 3D space to capture local geometric structures with higher precision. Complementarily, GeoSVR \cite{li2025geosvr} refines sparse voxel representations by enforcing geometry-aware regularization, which improves surface continuity and global consistency. In our work, we employ one of these approaches to reconstruct a high-quality triangular mesh from the GS representation, which we denote as  
\[
\mathcal{M} = (V, E, F),
\]  
where $V = \{v_i \in \mathbb{R}^3\}_{i=1}^{N_v}$ is the set of vertices, $E \subseteq V \times V$ is the set of edges, and $F \subseteq V \times V \times V$ represents the triangular faces. This mesh serves as a geometry-aware proxy that can be directly integrated into our VR framework.

\paragraph{Segmentation}
To enable object-level manipulation within VR environments, it is necessary to perform segmentation of the GS representation. In our system, we adopt a strategy similar to that employed in VR-GS~\cite{jiang2024vr}, where segmentation is achieved by associating subsets of Gaussians with individual object instances. Each segmented region corresponds to a coherent group of Gaussians that can be independently manipulated, deformed, or subjected to physics-based interactions. This approach not only supports efficient rendering and editing but also ensures that interactions are localized and consistent with object boundaries, thereby enhancing realism in immersive VR scenes. Alternatively, the Gaussians can be segmented as part of post-processing in cases where objects are visibly separated. 

\paragraph{GS-based generative models} 

Beyond reconstruction, GS also provides a foundation for generative 3D object synthesis, which is particularly useful for populating interactive VR environments. The TRELLIS framework~\cite{xiang2025structured} introduces a structured latent representation that can be decoded into multiple 3D formats, including 3D Gaussians and explicit meshes. 
TRELLIS generates small objects with both appearance and geometric detail. In our VR system, these generated mesh-augmented GS objects are employed as interactive props or scene elements, allowing users to grasp, manipulate, or collide with them while maintaining consistency with the underlying Gaussian representation. Formally, the generative model samples a latent $z \sim p(z)$ and decodes it into a Gaussian Splatting plus mesh pair $(\mathcal{G}, \mathcal{M})$, where $\mathcal{M} = (V, E, F)$ is the triangular mesh and $\mathcal{G}$ is the associated set of Gaussians. This dual output supports both efficient rendering via GS and geometry-aware interactions in VR.

\paragraph{\our{}: Mesh-based Gaussian Splatting representation}

\begin{figure}[t]
    \centering
    \centering
    \begin{tabular}{c c c c c}
         &  \multicolumn{4}{c}{Example modification}\\
        \raisebox{2\normalbaselineskip}[0pt][0pt]{\rotatebox{90}{Lamp}} & 
        \includegraphics[width=0.2\textwidth]{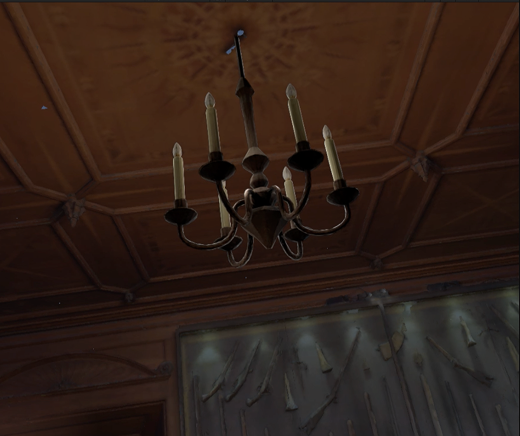} & 
        \includegraphics[width=0.2\textwidth]{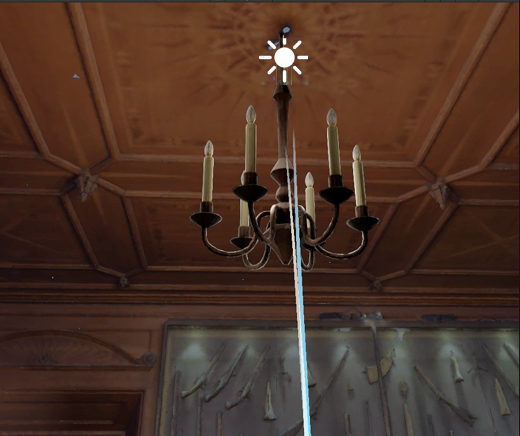} & 
        \includegraphics[width=0.2\textwidth]{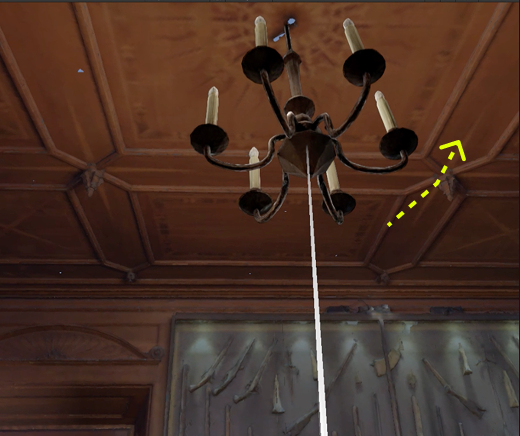} & 
        \includegraphics[width=0.2\textwidth]{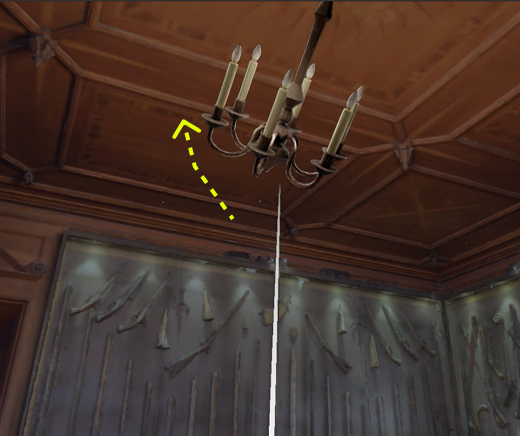} \\
        
        \raisebox{2\normalbaselineskip}[0pt][0pt]{\rotatebox{90}{Fox}} & 
        \includegraphics[width=0.2\textwidth]{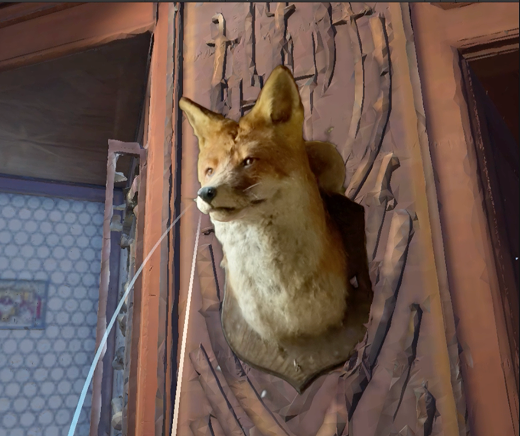} & 
        \includegraphics[width=0.2\textwidth]{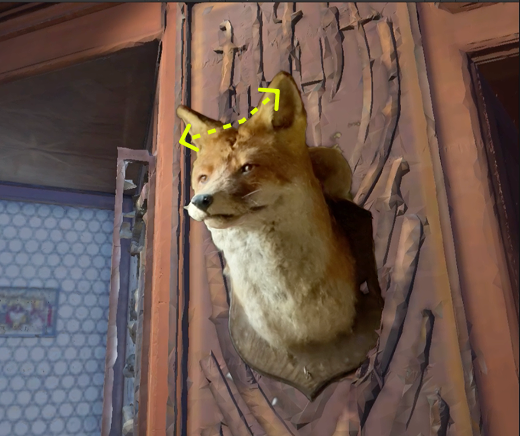} & 
        \includegraphics[width=0.2\textwidth]{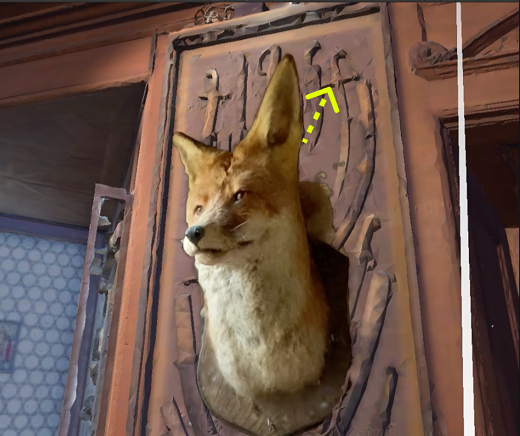} & 
        \includegraphics[width=0.2\textwidth]{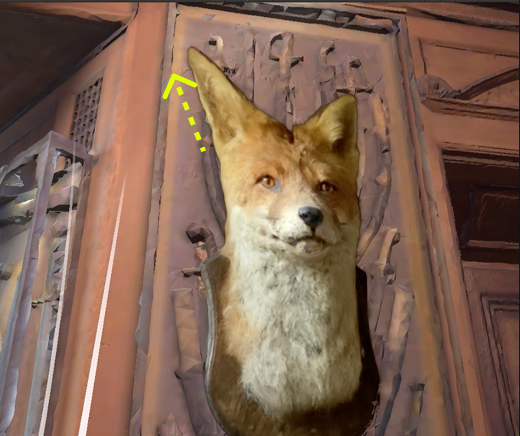}
    \end{tabular}
    \caption{Visual example of a participant performing the manipulations requested during the second, goal-directed action task: (1) \textit{Turn on the light by pointing at the lamp and swinging it to make it oscillate}, and (2) \textit{Pull the fox by the ears}.}
    \label{fig:darkroom}
\end{figure}

Our model (see Fig.~\ref{fig:method}) builds upon external mesh reconstruction tools to obtain an explicit triangular mesh representation, which subsequently serves as the basis for training Gaussian components. Following the approach of GaMeS~\cite{waczynska2024games}, we introduce a mesh-guided Gaussian parameterization, where splats are directly anchored to the mesh faces. Given a single triangular face with vertices 
$V = \{ {\bf v}_1, {\bf v}_2, {\bf v}_3 \} \subset \mathbb{R}^3$, 
the mean of a Gaussian component is defined as a convex combination of the vertices:  
\[
m_{V}(\alpha_1,\alpha_2,\alpha_3) = \alpha_1 {\bf v}_1 + \alpha_2 {\bf v}_2 + \alpha_3 {\bf v}_3,
\]  
where $\alpha_1+\alpha_2+\alpha_3=1$ and $\alpha_i$ are trainable parameters. This guarantees that the Gaussian centers remain consistently located within the face interior. To model the covariance, we construct a rotation matrix $R_V$ aligned with the triangle (using its normal and edge directions) and a diagonal scaling matrix $S_V$ proportional to the face dimensions. The covariance is then expressed as: 
\[
\Sigma_V = R_V S_V S_V^T R_V^T,
\]  
ensuring that each Gaussian adapts to the local face geometry. For a given face $V$, we place $k \in \mathbb{N}$ Gaussians, parameterized as:  
\[
\mathcal{G}_{V} = \left\{ \mathcal{N}\!\left(m_V(\alpha^{i}_1,\alpha^{i}_2,\alpha^{i}_3), \rho^{i}\Sigma_V\right) \right\}_{i=1}^k,
\]  
where $\rho^i \in \mathbb{R}_{+}$ controls the trainable scale. This parameterization tightly couples Gaussians with the underlying mesh, ensuring that geometric transformations of the mesh are directly propagated to the associated Gaussians. As a result, our representation enables consistent rendering and physical interaction, while preserving the structural properties of the original mesh, see Fig.~\ref{fig:vrgs_comparison}.  

\begin{figure}[t]
    \centering
    \centering
    \begin{tabular}{c c c c c}
         &  \multicolumn{4}{c}{Example modification}\\
        \raisebox{2\normalbaselineskip}[0pt][0pt]{\rotatebox{90}{Pillow}} & 
        \includegraphics[width=0.2\textwidth]{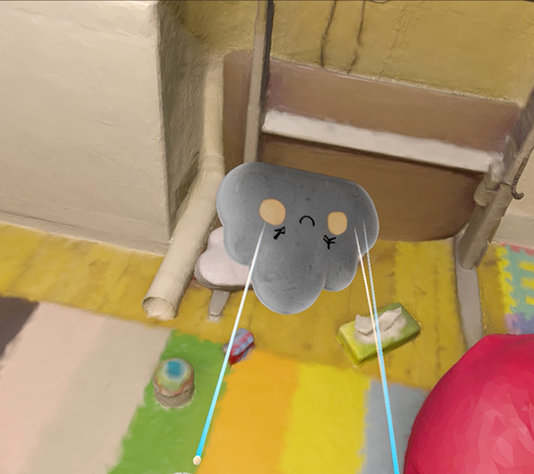} & 
        \includegraphics[width=0.2\textwidth]{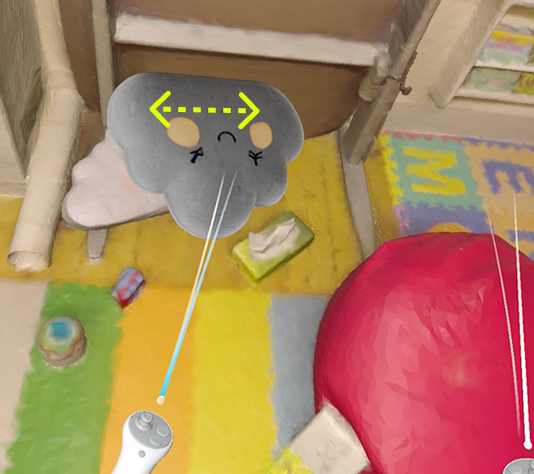} & 
        \includegraphics[width=0.2\textwidth]{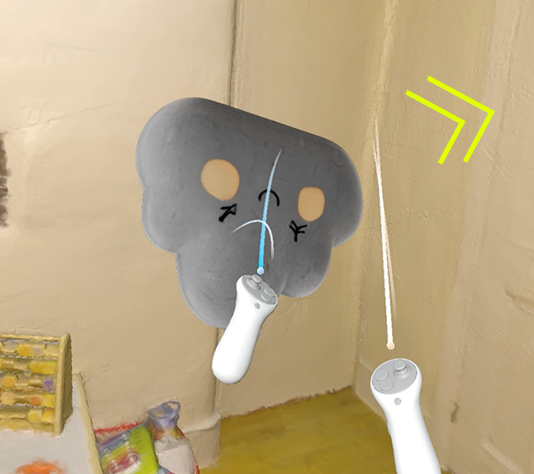} & 
        \includegraphics[width=0.2\textwidth]{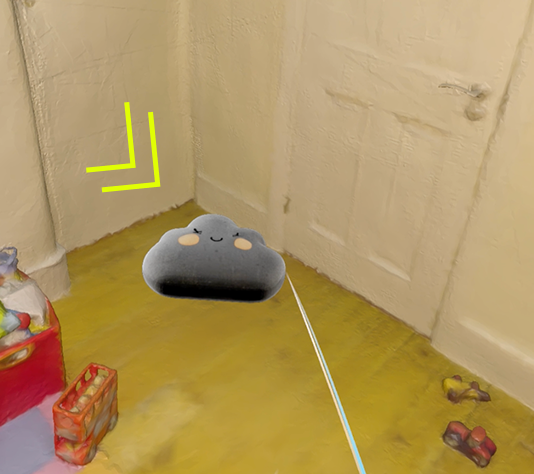} \\
        
        \raisebox{2\normalbaselineskip}[0pt][0pt]{\rotatebox{90}{Ballon}} & 
        \includegraphics[width=0.2\textwidth]{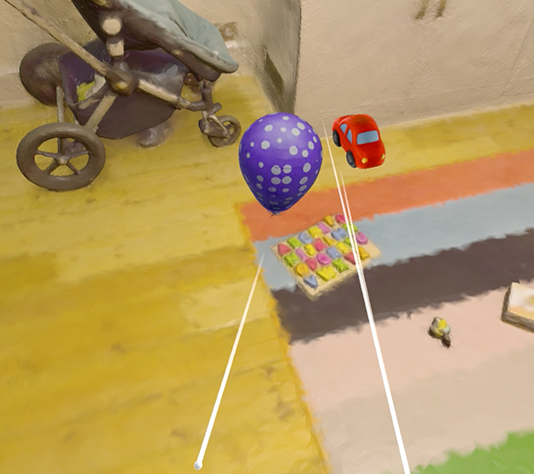} & 
        \includegraphics[width=0.2\textwidth]{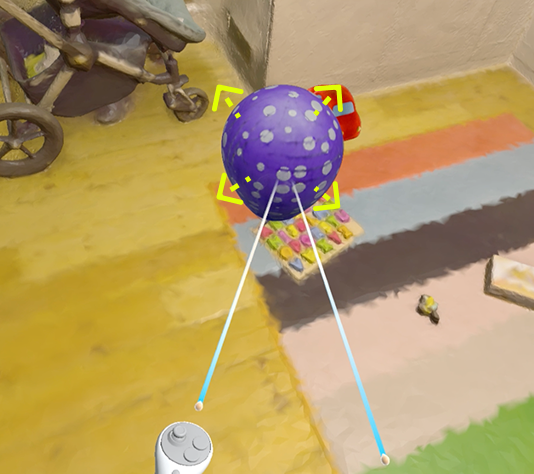} & 
        \includegraphics[width=0.2\textwidth]{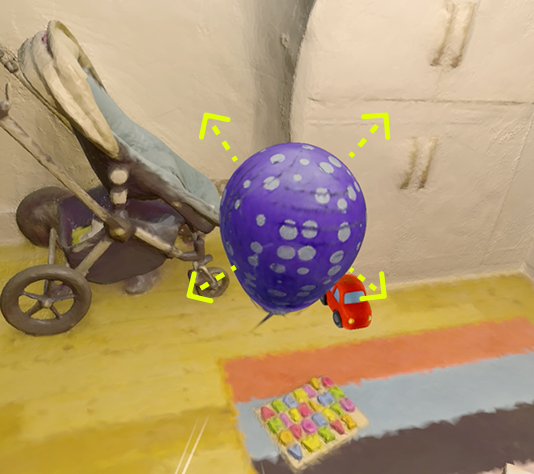} & 
        \includegraphics[width=0.2\textwidth]{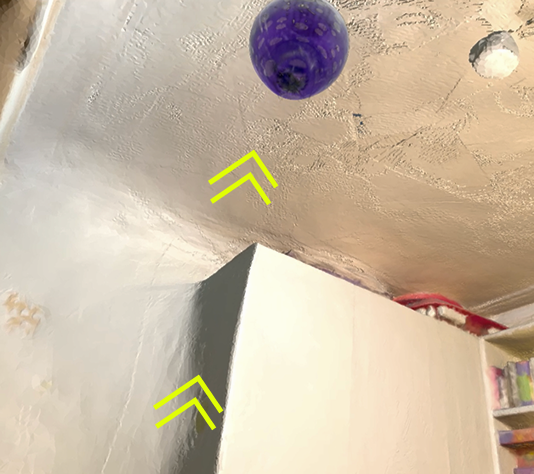}
    \end{tabular}
    \caption{Visual example of a participant performing the requested manipulations during the second, goal-directed action task: (1) \textit{Shake the pillow and throw it}, and (2) \textit{Inflate the balloon so that it rises}.} 
    \label{fig:toyroom}
\end{figure}

\paragraph{Physical simulations}  
A key advantage of our mesh-based GS representation is that it enables direct integration with existing physics simulation engines. Since our framework explicitly parametrizes Gaussians on top of a triangular mesh $\mathcal{M} = (V, E, F)$, we can leverage the vast ecosystem of physics solvers developed for mesh-based geometry. In practice, this enables us to employ standard techniques, such as mass-spring systems \cite{liu2013fast}, finite element methods \cite{marinkovic2019survey}, or position-based dynamics \cite{muller2007position}, to simulate deformations, collisions, and rigid-body interactions in a VR environment. Unlike approaches that approximate physical behavior through simplified proxies (e.g., low-resolution cages), our method ensures that physical forces act directly on the mesh, thereby indirectly propagating to the Gaussian splats anchored on its faces. This tight coupling provides physically consistent object behavior while maintaining the rendering efficiency of GS. As a result, our system achieves immersive and interactive dynamics in VR without requiring the development of specialized physics models.

\begin{figure}[t]
    \centering
    \centering
    \begin{tabular}{c c c c c}
         &  \multicolumn{4}{c}{Example modification}\\
        \raisebox{2\normalbaselineskip}[0pt][0pt]{\rotatebox{90}{Balls}} & 
        \includegraphics[width=0.2\textwidth]{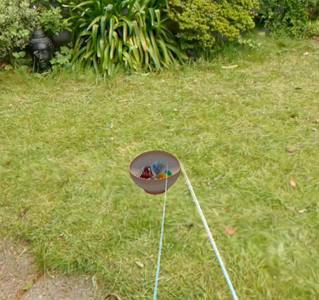} & 
        \includegraphics[width=0.2\textwidth]{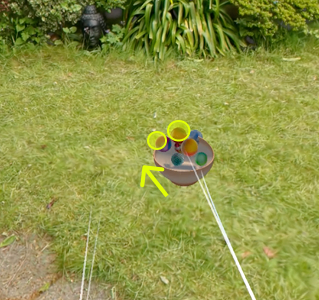} & 
        \includegraphics[width=0.2\textwidth]{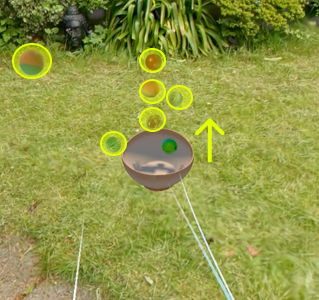} & 
        \includegraphics[width=0.2\textwidth]{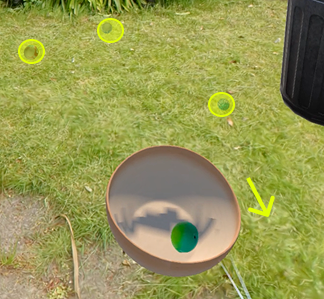} \\
        
        \raisebox{2\normalbaselineskip}[0pt][0pt]{\rotatebox{90}{Can}} & 
        \includegraphics[width=0.2\textwidth]{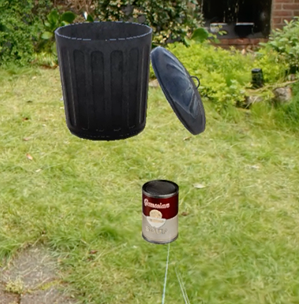} & 
        \includegraphics[width=0.2\textwidth]{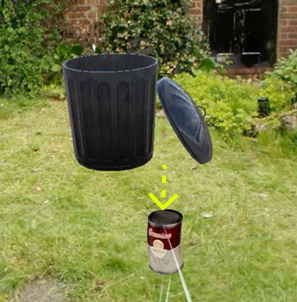} & 
        \includegraphics[width=0.2\textwidth]{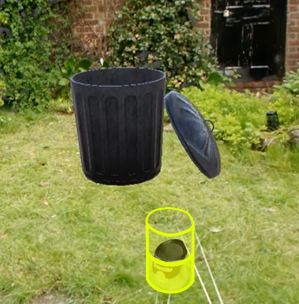} & 
        \includegraphics[width=0.2\textwidth]{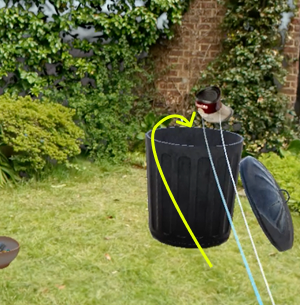}
    \end{tabular}
   \caption{Visual example of a participant performing the instructed manipulations requested during the second, goal-directed action task: (1) \textit{Tip the marbles out of the bowl}, and (2) \textit{Crush the can and throw it into the trash can}. The marbles were simulated using Unity rigid bodies with a physics material, while the can was simulated with a custom mesh deformation driven by applied forces.}
    \label{fig:gardenexamples}
\end{figure}

\section{Evaluation}

To evaluate our interface, we designed a user study with non-expert participants, as \our{} method should be used as a generalized tool allowing for generating various 3D assets that can be used to populate VR environments. When preparing our experimental design and tasks, we considered the prior evaluation approach that relied on user studies~\cite{jiang2024vr,franke2025vr,tu2025vrsplat}. During each task, we collected a range of subjective data to assess the perceived quality of rendered 3D assets. This included usability, measured with the \textit{System Usability Scale} (SUS) \cite{Brooke1996}, cognitive task load, measured with the \textit{NASA Task Load Index} (TLX) \cite{hart_1988_development}, and flow, measured with the \textit{Short Flow Scale} (SFS) \cite{engeser_flow_2Sanfeliu008}. We also administered the \textit{Simulation Sickness Questionnaire} (SSQ)  \cite{Kennedy01071993} to exclude participants experiencing ``severe' symptoms before using VR. These standardized questionnaires were presented to participants in their native language, using available translations from the literature. Finally, we used a Likert-like scale to ask participants how natural they rate the reconstructions generated by our method. All the statistical tests relied on a conservative $\alpha=0.05$ level.

\subsection{Experimental Setup}
We deployed the GS systems on a desktop PC equipped with Intel(R) Core(TM) i7-14700K (3.40 GHz), 32,0 GB RAM, NVIDIA GeForce RTX 4070 SUPER, running under Windows 11 Home OS and featuring CUDA compilation tools, release 12.4, V12.4.99 Build cuda\_12.4.r12.4/compiler.33961263\_0.
As the VR platform, we selected the Meta Quest Pro headset. For interacting with the VR environment and the GS models, we relied on Quest handheld controllers. The exact controls are shown in Tab.~\ref{tab:vr_buttons}.

\begin{table}[h]
\small
\centering
\begin{tabular}{p{3cm} p{5cm} p{5cm}}
\hline
\textbf{Button} & \textbf{Function Description} & \textbf{Example} \\
\hline
Primary Button & Used for \textit{stretch deformation} of virtual objects. When pressed and held, participants could stretch and elongate deformable objects to interact with their shape. & ‘Pull the fox by the ears’ (Task 2, \textit{dark room}, see Fig.~\ref{fig:darkroom}). \\[6pt]
Grab Button & Used for \textit{object grasping and manipulation}. Participants could grab, lift, and reposition objects within the virtual environment. & ‘Tip the marbles out of the bowl’ (Task 2, \textit{garden}, see Fig.~\ref{fig:gardenexamples}).  \\[6pt]
Select Button & Used for \textit{scaling, pressing, and adding movement}. This button allowed participants to scale object size, apply pressure, or introduce movement, depending on the task context. & ‘Inflate the balloon so
that it rises’(Task 2, \textit{toy room}, see Fig.~\ref{fig:toyroom}). \\
\end{tabular}
\caption{The mapping of the handheld controller button to manipulation/interaction methods.}
\label{tab:vr_buttons}
\end{table}

\subsection{3D Scenes}
For our evaluation study, we prepared three immersive scenes. Each of these scenes consisted of a large background model and three GS-based models generated with \our{} that could be manipulated and interacted with by the study participants. Furthermore, to ensure plausible manipulation results, the interaction with all these models was physics-aware, meaning that \textit{stretching}, \textit{twisting}, or \textit{shaking} closely resembled real-life manipulations. 

\paragraph{Scene: dark room}
We created this scene using the \textit{Armoury} 3D model from the \textit{Real World Textured Things} (RWTT) dataset \cite{Maggiordomo_2020}. The first interactable object, a \textit{fox}, was reconstructed from the \textit{Instant-NGP} dataset \cite{muller2022instant}. We implemented soft-body logic in Unity to simulate realistic deformation and dynamic movement in response to controller input. For the second and third objects, a \textit{chair} and a \textit{lamp}, we used TRELLIS to obtain the 3D meshes, and then Blender to generate the NeRF Synthetic dataset. The chair was configured as a rigid body with an assigned physics material and equipped with the \textit{XR Grab Interactable} component from the XR Interaction Toolkit, allowing it to be physically manipulated and moved by the user in virtual space. For the lamp, we used the \textit{XR Simple Interactable} component to listen for controller events and apply corresponding transformations for oscillation and turning the light on. The scene is shown in Fig. \ref{fig:darkroom}.

\paragraph{Scene: toy room}
We prepared this scene using the \textit{Playroom} scene from the Deep Blending dataset \cite{deepblending}. The images were used to create the coloured mesh using the mesh processing tool MeshLab \cite{meshlab}. The three interactable objects included: a \textit{balloon}, a \textit{car}, and a \textit{pillow}. These objects were also reconstructed using TRELLIS to generate the 3D meshes, followed by Blender to prepare the dataset. For the balloon, we used a Unity \textit{Compute shader} to simulate inflation and configured it as a rigid body with upward force so that it naturally rises. For the car, we configured it as a rigid body and added the \textit{XR Grab Interactable} component to listen for controller events and apply translation/grab transformations in response. The pillow was implemented as a soft body with logic to simulate stretching and deformation. In addition, it was equipped with the \textit{XR Grab Interactable} component to allow grabbing and throwing. The scene is shown in Fig. \ref{fig:toyroom}.

\paragraph{Scene: garden}
This scene was reconstructed using the \textit{Mip-NeRF 360 dataset} \cite{barron2022mip}. For the first of three interactable objects, \textit{marbles inside a bowl}, we prepared the dataset using TRELLIS and Blender, and each marble was configured as a rigid body with a physics material that caused them to bounce and jump naturally. The second object, a \textit{flower}, was segmented from the original garden scene and given the \textit{XR Grab Interactable} component to enable grabbing and throwing interactions. The last object, a \textit{can}, was reconstructed using TRELLIS and Blender and designed as a soft body with logic to simulate deformation under press forces applied via raycast. The scene is shown in Fig. \ref{fig:gardenexamples}.

\begin{table}[t]
\small
\centering
\setlength{\tabcolsep}{5pt}
\renewcommand{\arraystretch}{1.2}

\begin{tabular}{l@{\;\;}c@{\;\;}c@{\;\;}c@{\;\;}c@{\;\;}c@{\;\;}c@{\;\;}c}
\hline
\textbf{Manipulation} & \textbf{\makecell{Median\\VR-GS}} & \textbf{\makecell{Median\\\our{}} }& \textbf{\makecell{Mean $\pm$ SD\\VR-GS}} & \textbf{\makecell{Mean $\pm$ SD\\\our{}}} &\textbf{ W (Wilcoxon)} & \textbf{p-value }& \textbf{Effect size (r) }\\
\hline
\textit{stretching} & -0.5 & 2.0 & -0.33 $\pm$ 2.00 & 1.33 $\pm$ 1.28 & 4.0 & 0.009 & 0.62 \\
\textit{twisting}   & 0.5  & 1.0 & -0.28 $\pm$ 2.19 & 0.22 $\pm$ 1.90 & 29.0 & 0.425 & 0.19 \\
\textit{shaking}   & 2.0  & 2.0 & 1.50 $\pm$ 1.47 & 1.33 $\pm$ 1.61 & 15.0 & 0.673 & 0.10 \\
\end{tabular}

\caption{The descriptive statistics and Wilcoxon Signed-Ranks Test results comparing the perceived ``naturalness'' VR-GS and \our{} across three manipulations conducted during the first, close-ended task. The results suggest statistically significant difference for the \textit{stretching} manipulation between the two methods.}
\label{tab:wilcoxon_median}
\end{table}

\begin{table}[t]
\small
\centering
\setlength{\tabcolsep}{6pt}
\renewcommand{\arraystretch}{1.2}

\begin{tabular}{l@{\;\;}l@{\;\;}c@{\;\;}c@{\;\;}c@{\;\;}c@{\;\;}c@{\;\;}c}
\hline
\textbf{Questionnaire} & \textbf{\makecell{Median\\VR-GS}} & \textbf{\makecell{Median\\\our{}} }& \textbf{\makecell{Mean $\pm$ SD\\VR-GS}} & \textbf{\makecell{Mean $\pm$ SD\\\our{}} }& \textbf{W (Wilcoxon)} & \textbf{p-value} & \textbf{Effect size (r) }\\
\hline
SUS $\uparrow$ & 77.50 & 77.50 & 73.6 $\pm$ 16.7 & 78.9 $\pm$ 11.6  & 68.0 & 1.0000 & -0.000 \\
TLX $\downarrow$ & 7.085 & 6.665 & 9.81 $\pm$ 10.1 & 8.47 $\pm$ 6.12  & 55.0 & 0.5012 & 0.159 \\
\end{tabular}

\caption{Wilcoxon Signed-Ranks Test results comparing VR-GS and \our{} across usability and induced cognitive load during the first, close-ended task. The higher SUS [0-100] and lower TLX [0-100] are better. While the differences are statistically insignificant, our \our{} method led to higher SUS and lower TLX.}
\label{tab:wilcoxon_median_SUS_TLS_task1}
\end{table}

\begin{table}[t]
\small
\centering
\setlength{\tabcolsep}{6pt}
\renewcommand{\arraystretch}{1.2}

\begin{tabular}{l@{\;\;}c@{\;\;}c@{\;\;}c@{\;\;}c@{\;\;}c@{\;\;}c@{\;\;}c@{\;\;}c@{\;\;}c}
\hline
\textbf{Questionnaire} & 
\textbf{t} & 
\textbf{df} & 
\textbf{p-value} & 
\makecell{\textbf{Mean $\pm$ SD}\\\textbf{VR-GS}} & 
\makecell{\textbf{Mean $\pm$ SD}\\\textbf{\our{}}} & 
\makecell{\textbf{Mean}\\\textbf{difference}} & 
\makecell{\textbf{95\% CI}\\\textbf{lower}} & 
\makecell{\textbf{95\% CI}\\\textbf{upper}} & 
\textbf{Cohen's d} \\
\hline
FLOW  $\uparrow$ & -1.5 & 17 & 0.152 & 4.80 $\pm$ 1.07 & 5.03 $\pm$ 0.86 & -0.233 & -0.538 & 0.071 & -0.354 \\
\end{tabular}

\caption{Paired t-test comparing results of the flow experienced by the participants during the first, close-ended task execution using VR-GS and \our{}. Sample size (N), t-statistic, degrees of freedom (df), p-value, mean difference, 95\% confidence interval, and effect size (Cohen's d) are reported. The higher FLOW [1-7] values are better.}
\label{tab:flow_results}
\end{table}

\subsection{User Study}
\subsubsection{Participants}
We recruited 18 participants using opportunistic sampling. The group consisted of 2 females and 16 males, ranging in age from 19 to 31 years ($m=25.2(2)$, $SD=3.21$). 


\subsubsection{Experimental Design and Tasks}
 Inspired by earlier work \cite{jiang2024vr,tu2025vrsplat}, we designed two separate tasks to evaluate our approach with users. The first closed-ended task had a specific objective intended to compare our method with previous work concerning the use of GS in a VR environment \cite{jiang2024vr}. At the time this study was initiated, we selected VR-GS as the baseline for comparison because, to the best of our knowledge, it was the only publicly available implementation that combined Gaussian Splatting with physics-based object interaction in VR and provided example scenes, allowing for a direct and reproducible evaluation. Other concurrent works primarily focused on rendering efficiency or foveated visualization and did not offer publicly accessible implementations suitable for interactive assessment. Whereas the second, goal-directed action task served as a means of evaluating perceived quality of 3D scenes populated with 3D assets generated with our \our{} method. In both tasks, the manipulation with the 3D assets took into account physics-based constraints and affordances. Additionally, we administered the SSQ \cite{Kennedy01071993} questionnaire before each task to exclude participants who reported having ``severe'' symptoms before commencing any task.
 \begin{wrapfigure}{r}{0.5\textwidth}
\centering
\includegraphics[width=0.4\textwidth]{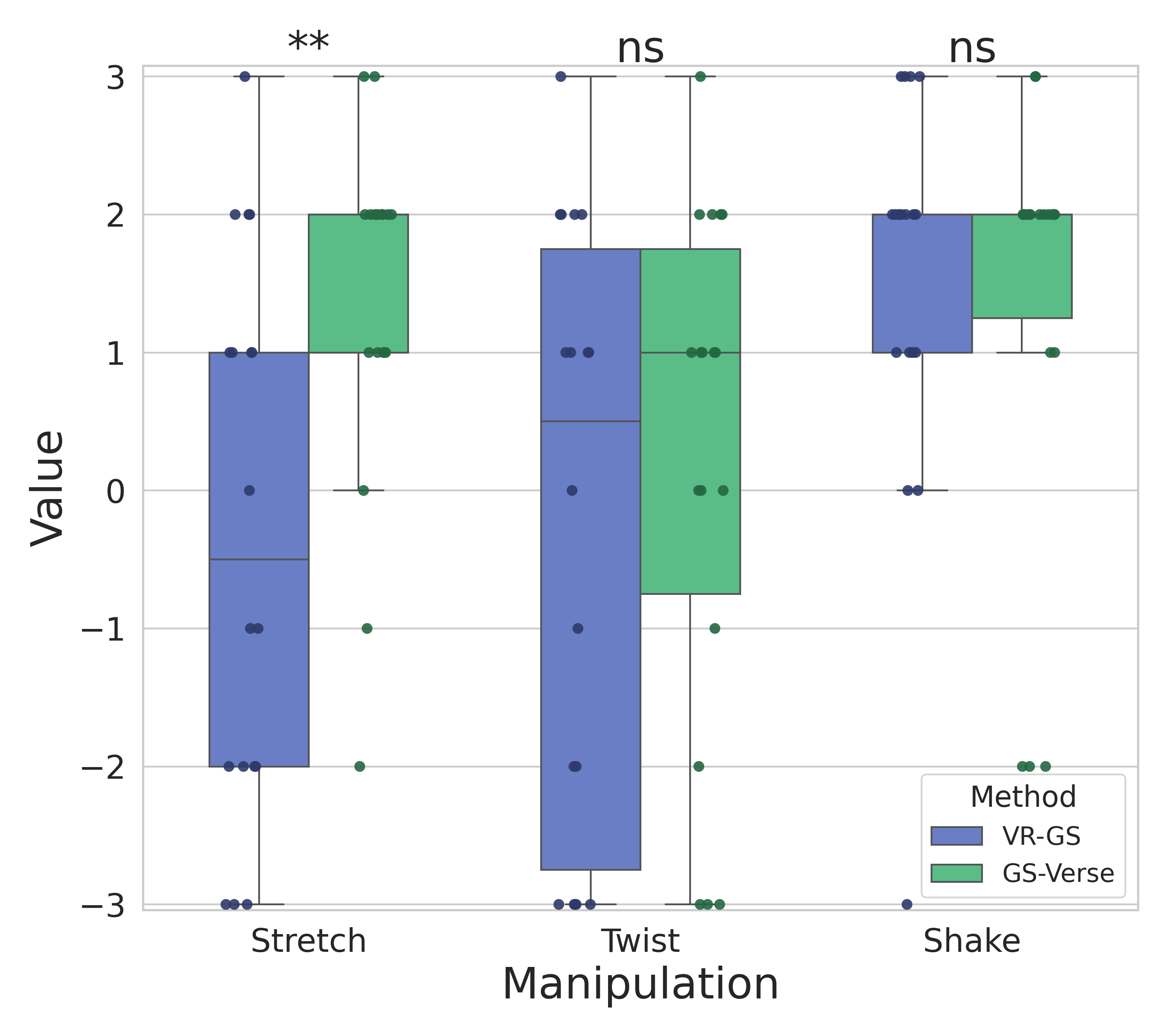}
\caption{Participants' rating of three physics-aware manipulations of the GS 3D objects: (i) \emph{stretching}, (ii) \emph{twisting}, and (iii) \emph{shaking} in task~1. The comparison revealed a statistically significant preference for our system in the \emph{stretching} manipulation, as well as higher consistency across the other two physics-aware manipulations (**p < 0.01, ns = not significant).}
\label{fig:task1_nature_comparison}
\end{wrapfigure}
 All participants initially experienced the test scene to familiarize themselves with available manipulation techniques. Afterward, they were exposed to the first closed-ended and second goal-directed action tasks. Such an approach allowed them to first learn about the available manipulation techniques (e.g., stretching, twisting, and shaking) and their limitations before freely experiencing the VR environment populated with 3D assets generated with \our{} method.

The first closed-ended task allowed us to directly compare \our{} with VR-GS \cite{jiang2024vr}. During this task, we asked the participants to (i) \textit{stretch}, (ii) \textit{twist}, and (iii) \textit{shake} a 3D object reconstructed in the same 3D scene by either \our{} or VR-GS presented in balanced, randomized order (see Fig.~\ref{fig:vrgs_comp_first_scene}). Participants were able to deform the 3D asset as much or as little as they wanted in under 2 minutes. After experiencing each of the two reconstruction methods, we administered the SUS \cite{Brooke1996}, TLX \cite{hart_1988_development}, and SFS \cite{engeser_flow_2Sanfeliu008} questionnaires. In addition, we asked the participants to assess the ``naturalness'' of the interaction with the 3D objects on a 7-point Likert scale ranging from \textit{non-natural} to \textit{very natural}.

\begin{figure}[t]
    \centering
    \begin{tabular}{c c c}
        SUS & TLX & FLOW \\
        \includegraphics[width=0.3\textwidth]{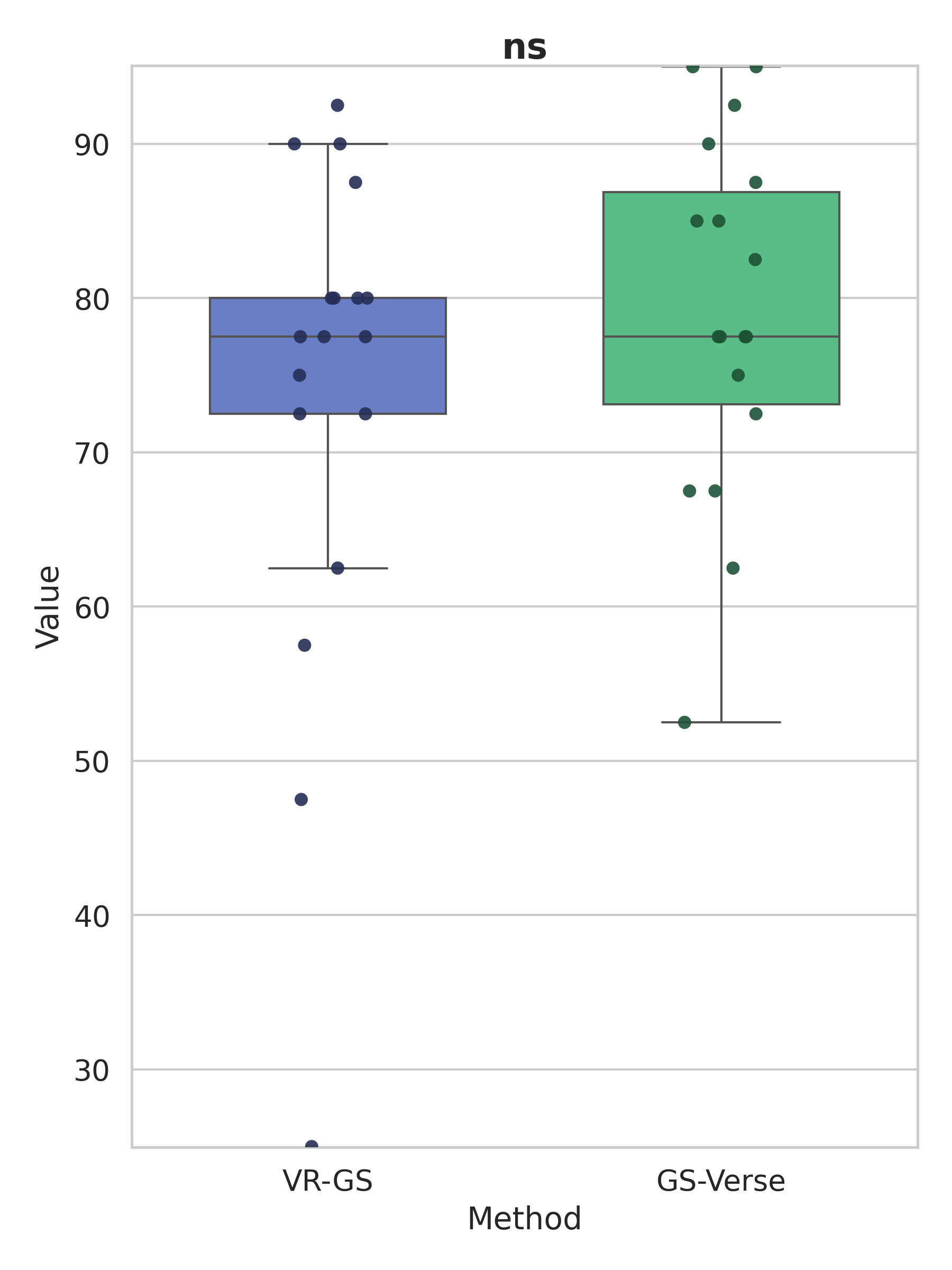} & 
        \includegraphics[width=0.3\textwidth]{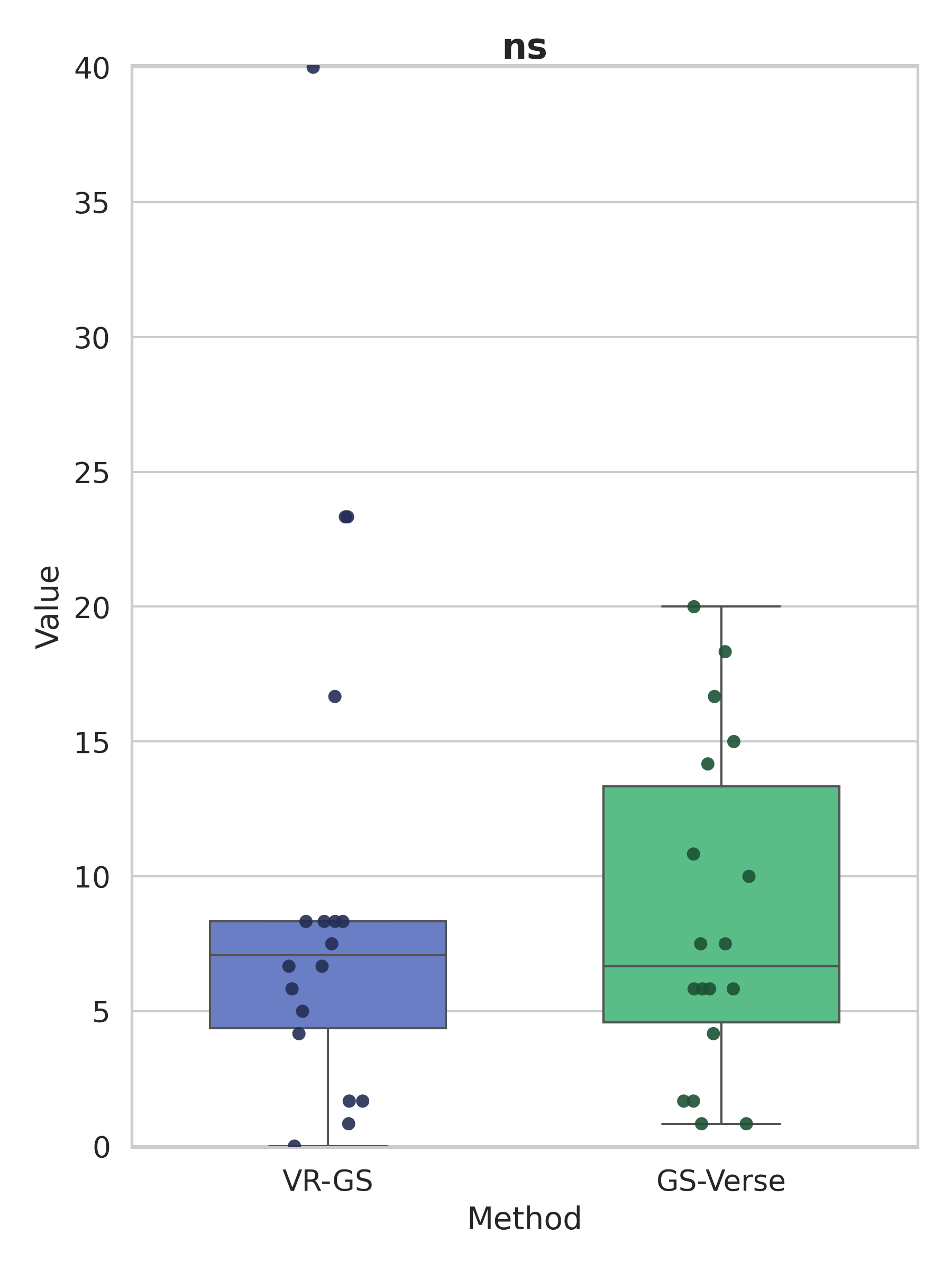} & 
        \includegraphics[width=0.3\textwidth]{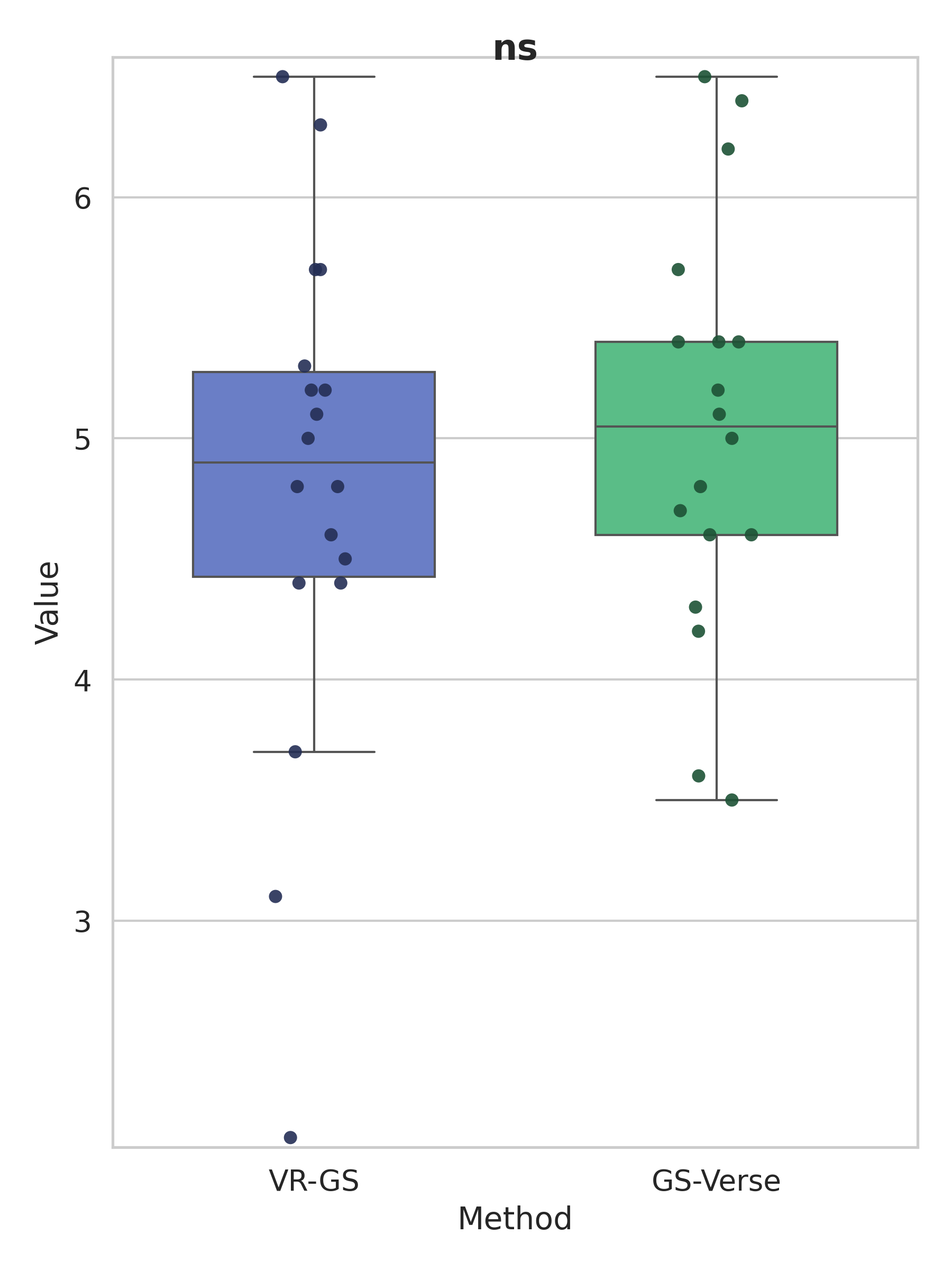}
    \end{tabular}
    \caption{Comparison of VR-GS and \our{} methods for the close-ended task across three questionnaires: SUS [0-100]$\uparrow$, TLX [0-100]$\downarrow$, and FLOW [1-7]$\uparrow$. Non-significance of paired comparisons between methods is indicated above each plot and calculated using the Wilcoxon Signed-Rank Test for SUS and TLX, and the paired t-test for FLOW (ns = not significant).}
    \label{fig:task1_plots_comparison}
\end{figure}

\begin{figure}[t]
    \centering
    \begin{tabular}{c c c}
        SUS & TLX & FLOW \\
        \includegraphics[width=0.3\textwidth]{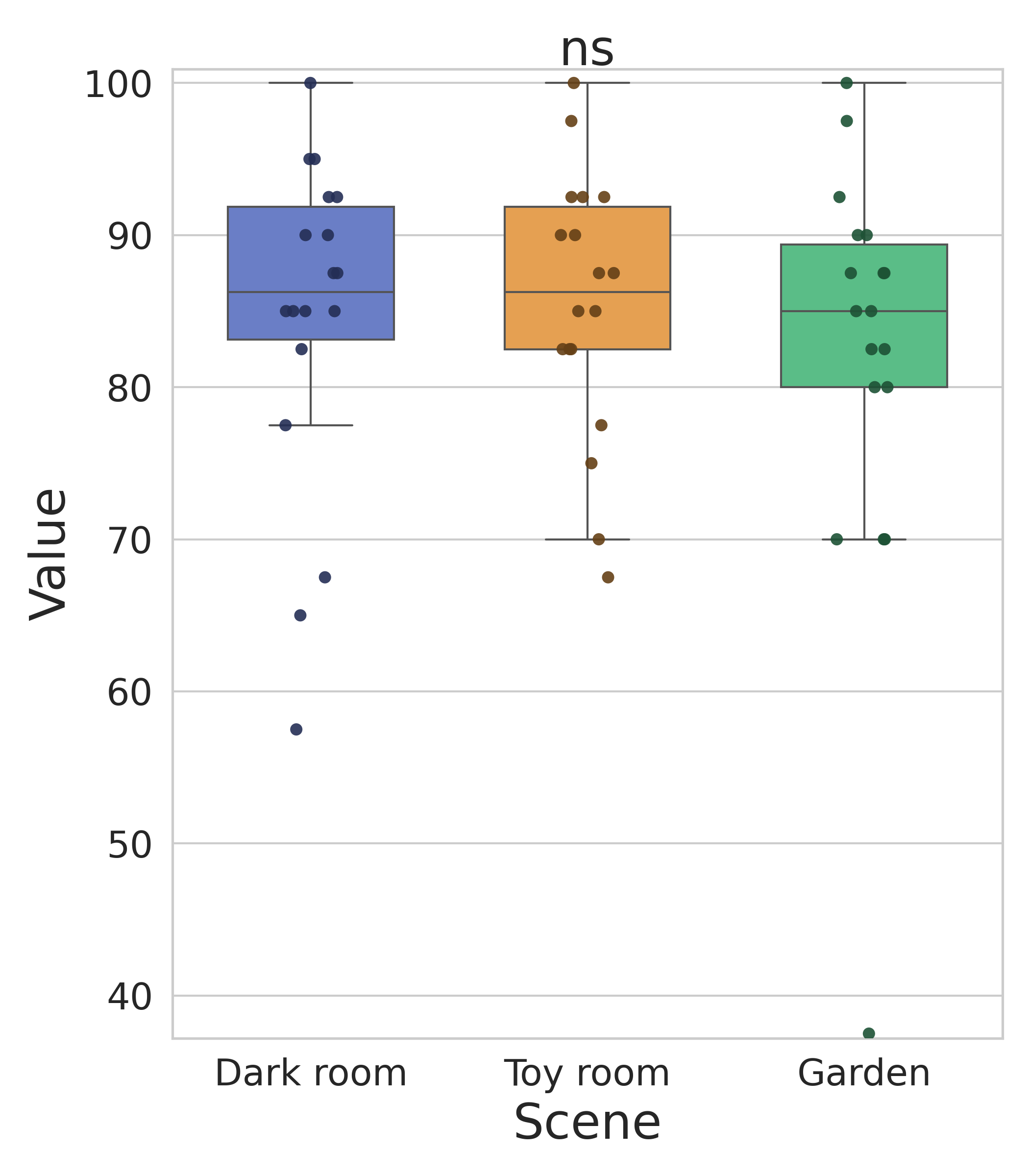} & 
        \includegraphics[width=0.3\textwidth]{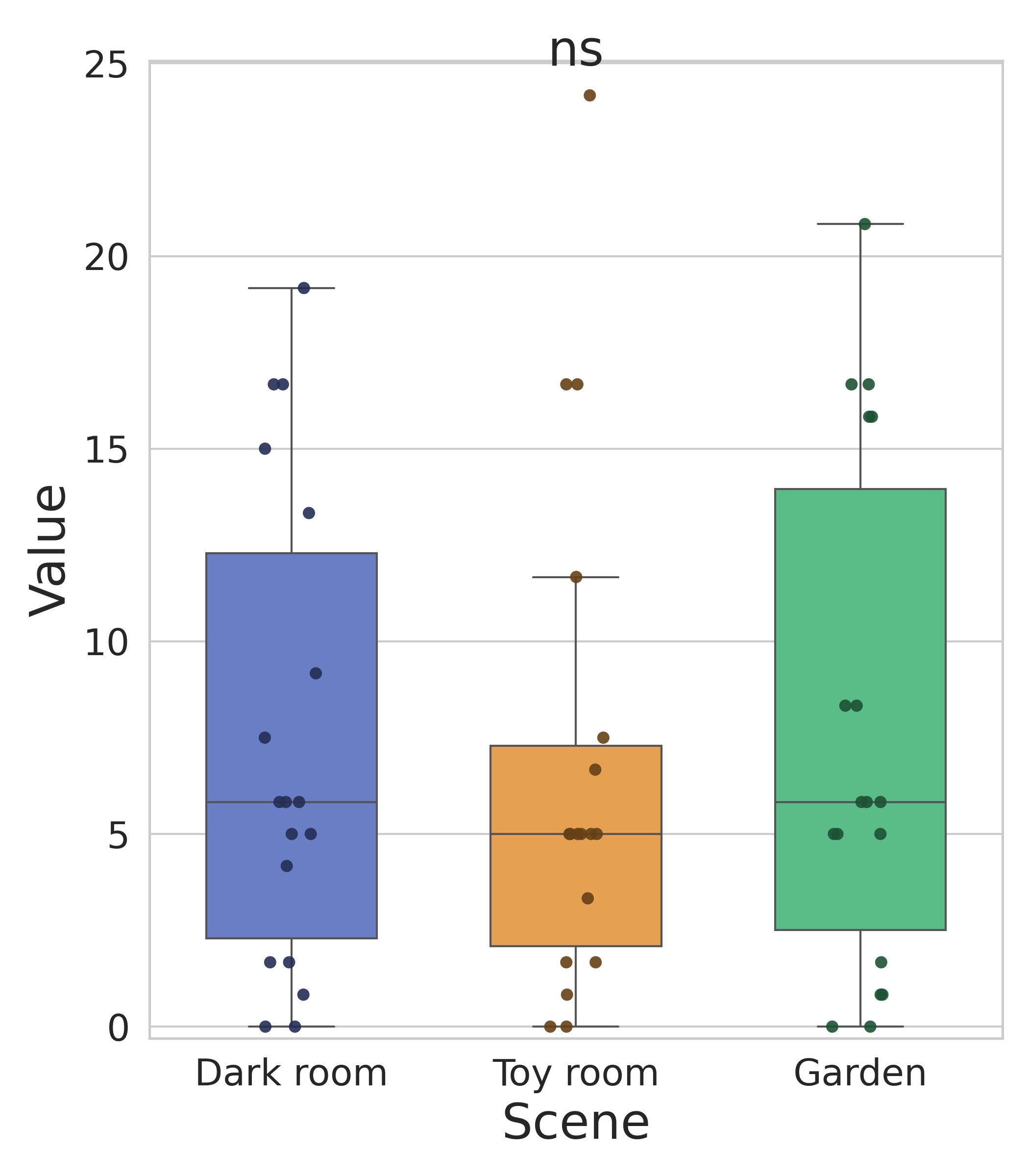} & 
        \includegraphics[width=0.3\textwidth]{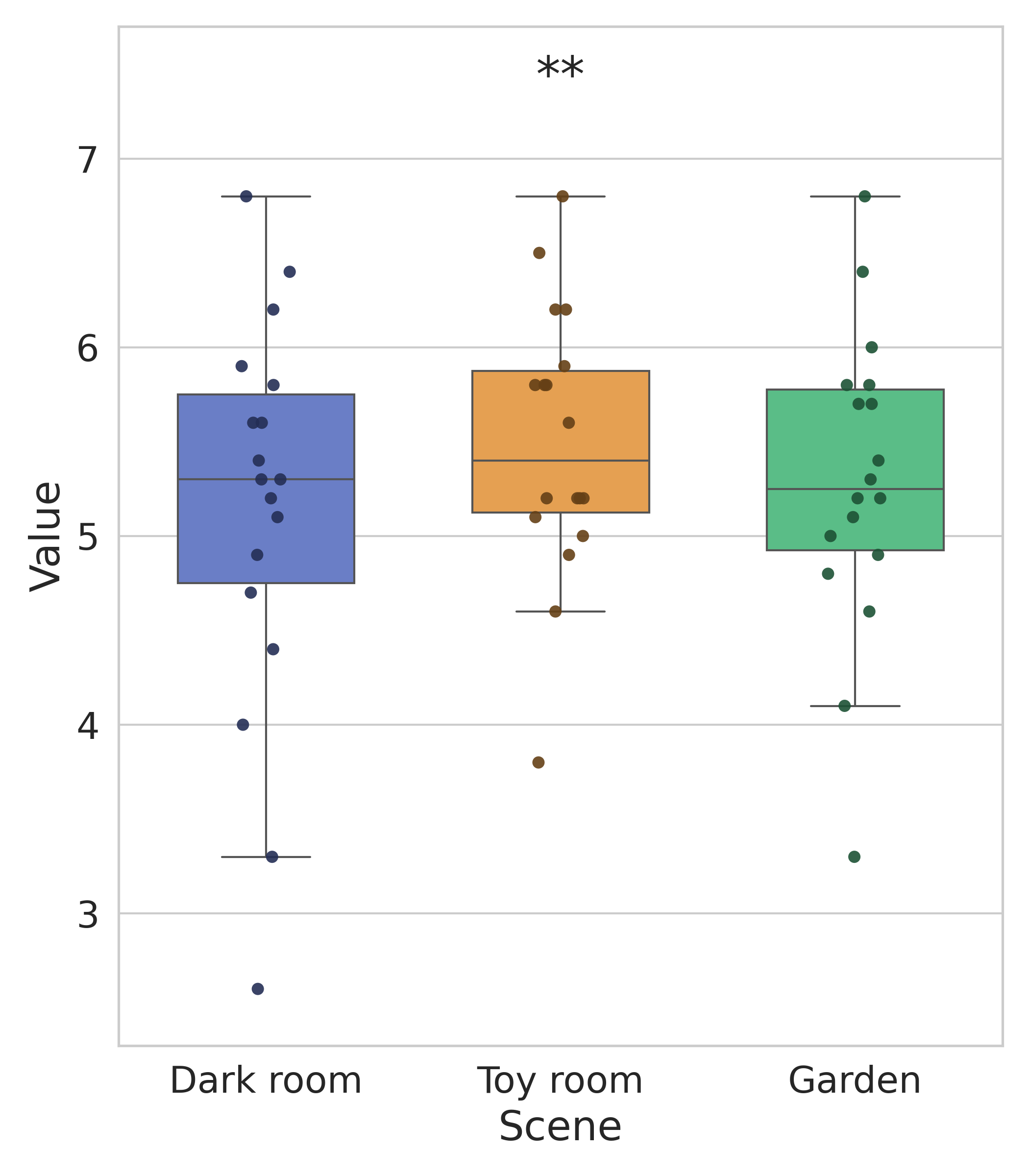}
    \end{tabular}
        \caption{Comparison of participants' provided scores for goal-directed action task across three evaluation methods: SUS [0-100]$\uparrow$, TLX [0-100]$\downarrow$, and FLOW [1-7]$\uparrow$. For SUS and TLX, a Friedman test was used to assess differences between conditions (scenes), while FLOW was analyzed using a repeated-measures ANOVA. Statistically significant difference was observed within the FLOW metric between the \textit{dark room} and the \textit{toy room} (**$p<0.01$, ns = not significant).}
    \label{fig:task2_SUS_TLX_SFS_comparison}
\end{figure}

The second task had a goal-directed action form and was also inspired by previous work coupling VR and GS \cite{tu2025vrsplat}. In this task, the participants were immersed in three different scenes, namely,  (A) \textit{dark room}, (B) \textit{toy room}, and (C) \textit{garden}. In each scene, we placed three interactive 3D assets that can be deformed using techniques available in an earlier task (see Fig. \ref{fig:teaser}). Furthermore, we asked the participant to perform three swift interactions, adjusted to be similar actions that could be taken in the particular real-life context and scene. In the case of the \textit{dark room}, these were: (1) \textit{Move the chair into the corner} (2) \textit{Turn on the light by pointing at the lamp, and swing it (or make it sway/oscillate)}, and (3) \textit{Pull the fox by the ears} (see Fig.~\ref{fig:darkroom}). When immersed in the \textit{toy room}, the participants were tasked to: (1) \textit{Inflate the balloon so that it rises}, (2) \textit{Start the toy car so that it crashes into the wall}, and (3) \textit{Shake the pillow and throw it} (see Fig.~\ref{fig:toyroom}). Whereas, when exposed to the \textit{garden} scene, we ask them to: (1) \textit{Tip the marbles out of the bowl}, (2) \textit{Knock the flowerpot off the table onto the grass}, and (3) \textit{Crush the can and throw it into the trash can} (see Fig.~\ref{fig:gardenexamples}). Similarly to the first task, after experiencing each of the three scenes, we administered the SUS \cite{Brooke1996}, TLX \cite{hart_1988_development}, and SFS \cite{engeser_flow_2Sanfeliu008} questionnaires. In addition, we asked the participants to assess the ``naturalness'' of the interaction with the 3D objects on a 7-point Likert scale ranging from \textit{non-natural} to \textit{very natural}. This approach allowed us to gain insight into how participants perceived the quality of the renders and physics-aware interaction provided with \our{} method under a diverse range of contexts.


\section{Evaluation Results}

To compare our \our{} method with prior work and assess the overall quality of our approach for populating interactive, immersive environments across various contexts, we employed a qualitative evaluation strategy. We decided on this evaluation because the primary goal of our work is to provide plausible, realistic, physics-aware manipulations of 3D assets within a VR environment populated with GS-based objects. Therefore, we employed a Likert-like scale to capture participants' own assessment of the ``naturalness'' of the three distinguishable, physics-based manipulations, as well as three standardized, well-established surveys. 

The latter included the SUS \cite{Brooke1996} questionnaire for assessing the perceived usability of interacting with and manipulating 3D assets generated with our \our{} method. We also administered the NASA-TLX questionnaire \cite{hart_1988_development}, a widely adopted approach for quantifying participants' subjective cognitive load during the execution of a specific task in VR \cite{Tadeja_3D_Photogrammetry_VR, tadeja_exploring_2023}. Finally, we used the SFS questionnaire \cite{engeser_flow_2Sanfeliu008} to track perceived flow levels during VR interaction, which can serve as an indicator of user engagement and skill during task performance \cite{engeser_flow_2Sanfeliu008, laakasuo_psychometric_2022_FLOW}. 

\subsection{Close-Ended Task}

\paragraph{Perceived Naturalness}
The Shapiro-Wilk test indicated that the data were not normally distributed. Therefore, we used the nonparametric Wilcoxon Signed-Rank Test. We present in Tab. \ref{tab:wilcoxon_median} results comparing medians and test statistics between VR-GS and \our{} (our) across three physics-aware manipulations, i.e., \textit{stretching}, (ii) \textit{twisting}, and (iii) \textit{shaking}. For the latter two, the Wilcoxon Signed-Rank Test revealed no statistically significant difference (\textit{twisting} $W = 29.00$, $p = .425$, $r = .19$; \textit{shaking}: $W = 15.00$, $p = .673$, $r = .10$) (see Fig.~\ref{fig:task1_nature_comparison}). However, in the case of \textit{stretching} manipulation, the median performance was higher for the \our{} method ($Mdn = 2.00$) than for VR-GS ($Mdn = –0.50$). A Wilcoxon Signed-Rank Test indicated that this difference was statistically significant, $W = 4.00$, $p = .009$, with a large effect size ($r = .62$).
This suggests that participants perceived the \textit{stretching} manipulation as significantly more natural when using our \our{}, with VR-GS showing visible artifacts such as large overlapping splats and detached parts (see Fig.~\ref{fig:beararteffact}). Moreover, as indicated in Fig.~\ref{fig:task1_nature_comparison}, they were also more consistent in their assessment of the remaining two manipulations.




\paragraph{Perceived Usability} 
For the SUS questionnaire, median performance was identical between VR-GS ($Mdn = 77.50$) and \our{} ($Mdn = 77.50$). On average, the VR-GS method scored $73.61 \pm 16.68$, while \our{} method scored $78.89 \pm 11.58$, which is above average in both cases, with our method achieving a slightly higher result than the overall mean for graphical user interfaces (GUIs) \cite{bangor_determining_2009}.
As the Shapiro-Wilk test indicated that the data were not normally distributed, we used the Wilcoxon Signed-Rank Test, which showed that the difference was not statistically significant ($W = 68.0$, $p = 1.000$), with a negligible effect size ($r = -0.000$) (see Tab.~ \ref{tab:wilcoxon_median_SUS_TLS_task1}). 

\paragraph{Perceived Taskload} 
With respect to TLX results, median workload scores were slightly higher for VR-GS ($Mdn = 7.085$) than for \our{} (Mdn = 6.665). The mean workload for VR-GS was $9.81 \pm 10.11$, compared to $8.47 \pm 6.13$ for \our{}.  
Again, due to the non-normality of the data, we relied on the nonparametric Wilcoxon Signed-Rank Test, which showed no statistically significant difference ($W = 55.0$, $p = 0.501$), with a small effect size ($r = 0.159$)(see Tab.~ \ref{tab:wilcoxon_median_SUS_TLS_task1}). 
The mean scores in both groups can be considered to belong to the ``low'' category \cite{Prabaswari_2019_NASA_TLX_Table, tadeja_exploring_2023}, as expected given the ease of the manipulation task and the highly usable interface, as shown by the high SUS scores.
 

\paragraph{Perceived Flow} 
For the FLOW method, participants’ scores were slightly lower for VR-GS ($M = 4.80$, $SD = 1.07$) compared to \our{} ($M = 5.03$, $SD = 0.86$). 
A paired-samples t-test indicated that this difference, $\bar{d} = -0.233$, 95\% CI [$-0.538$, $0.071$], was not statistically significant, $t(17) = -1.50$, $p = .152$, with a small effect size (Cohen's $d = -0.354$) (see Tab.~\ref{tab:flow_results}). 

\paragraph{Observed Manipulation Issues}
We summarize in Tab.~\ref{tab:stats_interaction_issues} the number and percentage of participants who experienced interaction issues in VR-GS and \our{}. Interaction latency issues refer to delays in the system's response to participant-invoked manipulations, while interaction range awareness issues indicate participants’ difficulty judging the effective interaction distance. In VR-GS, 5 out of 18 participants (27.8\%) reported latency delays during task performance, despite all users operating under the same conditions and hardware. Furthermore, 11 participants (61.1\%) experienced difficulties in accurately perceiving the interaction range from the outset, without guidance or hints. In contrast, with \our{}, only 3 participants (16.7\%) reported latency issues, and just 2 participants (11.1\%) faced interaction range awareness difficulties, indicating that our system provides a more stable and natural interaction experience.

\paragraph{Overall Assessment}
We observed no statistically significant differences between VR-GS and \our{} in terms of usability (SUS), taskload (TLX), and (FLOW) (see Fig.~\ref{fig:task1_plots_comparison}). However, our approach led to perceiving physics-aware manipulations as significantly more natural for \textit{stretching} 3D objects and to a more consistent assessment of other manipulations, such as \textit{twisting} and \textit{shaking} (see Fig.~\ref{fig:task1_nature_comparison}).

\begin{table*}[t]
\small
{\setlength{\tabcolsep}{8.7pt}
{\fontsize{8.5pt}{11.5pt}\selectfont
\begin{tabular}{lccc}
\hline
\bf Issue Type & \bf VR-GS (n = 18) & \bf \our{} (n = 18) \\ \hline
\textit{Interaction latency issue} $\downarrow$ & 5 (27.78\%) & 3 (16.67\%) \\
\textit{Interaction range awareness issue} $\downarrow$ & 11 (61.1\%) & 2 (11.1\%) \\
\end{tabular}
}
}
\caption{Number and percentage of participants experiencing interaction issues in VR-GS and \our{}. ``Interaction latency issues'' indicate moments of delayed system response, while ``interaction range awareness issues'' indicate participants’ difficulty judging the effective interaction distance.
}
\label{tab:stats_interaction_issues}
\end{table*}

\subsection{Goal-Directed Action Task} 

\paragraph{Perceived Usability} 
Participants’ usability scores were assessed across three scenes using the System Usability Scale (SUS). The median SUS scores were the same for the \textit{dark room} and \textit{toy room} (86.25), and slightly lower for the \textit{garden} scene (85.00). Because the Shapiro-Wilk test indicated that the data were not normally distributed, we used a nonparametric Friedman test to examine differences in SUS scores across the three scenes. The analysis showed no statistically significant difference, $\chi^2(2) = 2.81$, $p = 0.245$ (see Fig.~\ref{fig:task2_SUS_TLX_SFS_comparison}), while, at the same time, all the results are high and well above the mean for GUIs \cite{bangor_determining_2009}. 

\paragraph{Perceived Taskload} 
Median workload scores were also high and similar across the three scenes, though slightly lower for the \textit{toy room} scene, i.e., 5.83, 5.00, and 5.83, respectively. The mean taskload scores were $7.41 \pm 6.20$, $6.71 \pm 6.57$, and $7.68 \pm 6.63$ for \textit{dark room}, \textit{toy room}, and \textit{garden} scenes. As the data were non-normal, we conducted a Friedman test, which revealed no statistically significant differences across the rooms, $\chi^2(2) = 4.33$, $p = 0.115$ (see Tab.~\ref{tab:nasa_sus_flow_sidebyside_task2}). Similarly to the close-ended task, the mean scores fall within the ``low'' workload category \cite{Prabaswari_2019_NASA_TLX_Table, tadeja_exploring_2023}. 

\paragraph{Perceived Flow} 
For the FLOW method, the scores were slightly lower in the \textit{dark room} scene ($M = 5.14$, $SD = 1.06$) compared to the \textit{toy room} ($M = 5.49$, $SD = 0.73$) and \textit{garden} ($M = 5.28$, $SD = 0.81$). A one-way repeated-measures ANOVA indicated that these differences were statistically significant, $F(2, 34) = 5.47$, $p = .009$, $\eta_p^2 = .244$. Post hoc comparisons with Bonferroni correction showed that scores associated with the \textit{toy room} scene were significantly higher than in \textit{dark room}, $p = .024$, 95\% CI [$-$0.60, $-$0.10] (see Tab.~\ref{tab:task2_flow_pairwise}). While the differences between \textit{garden} and \textit{dark room} ($p = .602$, 95\% CI [$-$0.37, 0.08]) and \textit{toy room} ($p = .122$, 95\% CI [0.01, 0.40]) scenes were not statistically significant (see Tab.~\ref{tab:nasa_sus_flow_sidebyside_task2}).

A possible explanation for the observed statistical significance between \textit{dark room} and \textit{toy room} scenes is a disruption of the balance between task difficulty and participants’ skills. The flow state occurs when a task provides an optimal level of challenge, i.e., when it is engaging but not overly demanding \cite{Csikszentmihalyi91}. In the \textit{dark room} environment, reduced lighting, lower object visibility, and higher visual complexity may have made orientation and task performance more difficult, partially disturbing this balance. Additionally, the high number of visual details, combined with lower illumination, may have influenced this result, as the environment could be considered more perceptually demanding, potentially negatively affecting the intensity of the flow experience\cite{hassan2020flow}. Moreover, the lack of clearly defined goals in the \textit{dark room} condition—for instance, slightly ambiguous instructions such as ``move the chair to another corner of the room,'' without specifying which chair or which corner—may have further hindered participants’ sense of control and clarity, thereby potentially reducing the experienced flow.

\paragraph{Overall Assessment}
Our analysis revealed no statistically significant differences between scenes in terms of usability (SUS) or taskload (TLX). However, a statistically significant difference was found for the FLOW measure between the \textit{dark room} and \textit{toy room} scenes. This finding suggests that the \textit{dark room} environment may have disrupted the balance between task difficulty and participants’ skills. The \textit{dark room} environment’s reduced lighting, greater visual complexity, and less clearly defined goals likely increased perceptual and cognitive demands, potentially reducing the overall intensity of the flow experience.

        

\begin{table}[t]
\small
\centering
\begin{tabular}{@{}l@{\;}c@{\;}c|l@{\;}c@{\;}c|l@{\;}c@{\;}c@{}}
\multicolumn{3}{c}{\textbf{NASA TLX}} & \multicolumn{3}{c}{\textbf{SUS}} & \multicolumn{3}{c}{\textbf{FLOW}}  \\
\hline
Scene & Median & Mean $\pm$ SD & Scene & Median & Mean $\pm$ SD & Scene & Median & Mean $\pm$ SD \\
\hline
\textit{dark room} & 5.83 & 7.41 $\pm$ 6.20 & \textit{dark room} & 86.25 & 84.44 $\pm$ 11.17 & \textit{dark room} & 5.30 & 5.14 $\pm$ 1.06 \\
\textit{toy room}  & 5.00 & 6.71 $\pm$ 6.57 & \textit{toy room}  & 86.25 & 85.42 $\pm$ 8.80 & \textit{toy room} & 5.40 & 5.49 $\pm$ 0.73 \\
\textit{garden}    & 5.83 & 7.68 $\pm$ 6.63 & \textit{garden}    & 85.00 & 81.94 $\pm$ 14.00 & \textit{garden} & 5.25 & 5.28 $\pm$ 0.81 \\
\hline
\multicolumn{3}{c}{\textbf{Friedman Test}} & \multicolumn{3}{c|}{\textbf{Friedman Test}} & \multicolumn{3}{c}{\textbf{ANOVA Test}} \\
\hline
Chi-Square & p-value & df & Chi-Square & p-value & df & Mean Square & F & df \\
4.33 & 0.115 & 2 & 2.81 & 0.245 & 2 & 0.663 & 5.47 & 2 \\
\hline
\multicolumn{6}{c|}{} & Sig. & Partial $\eta^2$ & \\ 
\multicolumn{6}{c|}{} & 0.009 & 0.244 & \\ 
\end{tabular}
\caption{Descriptive statistics and results of the Friedman tests for NASA TLX (left) and SUS (middle), and results of the repeated-measures ANOVA for FLOW (right), across the three scenes — \textit{dark room}, \textit{toy room}, and \textit{garden} — used in the second, goal-directed action task. No statistically significant differences were observed for NASA TLX and SUS; however, significant differences were found for Flow.}
\label{tab:nasa_sus_flow_sidebyside_task2}
\end{table}

\begin{table}[t]
\small
\centering
\begin{tabular}{@{}l@{\;\;}c@{\;\;}c@{\;\;}c@{\;\;}c@{\;\;}c@{\;\;}c@{\;\;}c@{}}
\hline
\bf Comparison & \bf Mean Diff & \bf Std. Error & \bf t & \bf df & \bf p (Bonferroni) & \bf 95\% CI Lower & \bf 95\% CI Upper \\
\hline
\textit{dark room vs toy room} & -0.350 & 0.116 & -3.01 & 17 & 0.024 & -0.596 & -0.104 \\
\textit{dark room vs garden} & -0.144 & 0.109 & -1.33 & 17 & 0.602 & -0.373 & 0.085 \\
\textit{toy room vs garden} & 0.206 & 0.093 & 2.22 & 17 & 0.122 & 0.010 & 0.401 \\
\end{tabular}
\caption{The results of pairwise comparisons (with Bonferroni correction) of FLOW scores across three scenes, i.e., \textit{dark room}, \textit{toy room}, and \textit{garden}), used in the second, goal-directed action task. This outcome suggests a statistically significant difference in experienced flow between the \textit{dark room} and the \textit{toy room}.}
\label{tab:task2_flow_pairwise}
\end{table}

\section{Discussion}

We propose a unified, physics-aware interactive system that achieves real-time interactions with 3D GS representations in VR. Our approach was guided by three core principles: (i) achieving immersive, realistic generative dynamics; (ii) while ensuring high-quality and resource-efficient physics-aware manipulation across various contexts and use case scenarios; and (iii) offering a unified framework that links visual representation and physical simulation.

\paragraph{Immersive and Realistic Generative Dynamics} The primary goal of our system is to provide users with an immersive experience where virtual objects behave and deform in a manner that closely mirrors the real world \cite{kalawsky1999vruse}. This necessitates a move beyond purely geometric deformation towards a system grounded in physical principles. Unlike methods that reconstruct motion from time-dependent datasets or use generative machine learning, our system simulates dynamics based on physical laws.

A key limitation of a handful of existing prior works, such as VR-GS \cite{jiang2024vr}, is their reliance on a simplified tetrahedral mesh (cage) reconstructed from Gaussian kernels to drive the simulation. While computationally efficient, this cage only approximates the true object surface. These core characteristics may lead to object manipulations that users do not perceive as natural. As noted in the VR-GS, the occurrence of artifacts during large deformation necessitates a more robust approach than the current simple embedding strategy \cite{jiang2024vr}. To mitigate this, they introduced a more complex two-level interpolation scheme.


In contrast, our system, \our{}, addresses this problem at its root. By directly integrating the object's mesh with the GS representation, we eliminate the need for such approximations and corrective workarounds. This direct binding ensures that the physical forces are applied to a geometrically accurate surface, resulting in deformations that are not only physically plausible but also visually faithful to the object's intricate details. To evaluate our method, we carried out a comparative user study involving three different types of physics-aware object manipulations in VR. Our approach led to perceiving physics-aware manipulations as significantly more natural when \textit{stretching} 3D objects and to more consistent results for \textit{twisting} and \textit{shaking} (see Fig.~\ref{fig:task1_nature_comparison}). Further examination of our method in different scenes and contexts (see Fig. \ref{fig:teaser}) shows its high usability ($\geq85.0/100.0$), low taskload ($\leq10.0/100.0$), and high flow ($\geq5.0/7.0$) when manipulating 3D objects generated with our method and populating a VR environment (see Tab. \ref{tab:nasa_sus_flow_sidebyside_task2}).


\paragraph{Physics-Aware Resource-Efficient Manipulation} Efficient use of available computing resources is crucial for interactive VR systems to prevent user discomfort and simulation sickness \cite{deng2022fov,sutcliffe2019reflecting}. While fundamental transformations like rotation, scaling, and translation are standard, our system enables real-time, physics-based interactions that allow for direct manipulation and deformation of objects. The challenge lies in achieving this without exceeding VR's strict frame budget. The authors of VR-GS correctly identify that a per-Gaussian simulation, as seen in PhysGaussian \cite{xie2024physgaussian}, is computationally prohibitive for real-time applications. Their choice of a reduced tetrahedral mesh and an XPBD-based simulation was a pragmatic compromise to meet performance targets.

Our system builds on this insight but takes a different path to optimization. By leveraging the original mesh, we can employ well-established and highly optimized algorithms for mesh-based physics simulation. Crucially, we introduce a novel mapping algorithm that efficiently propagates deformations from mesh vertices to the thousands of associated Gaussian kernels in a single, parallel-friendly computational pass. This allows us to handle complex and accurate physical representation while still operating comfortably within the real-time frame rate, thus achieving the goal of physically realistic editing without the performance penalty of per-primitive calculations.

\paragraph{Unified Framework.} Our design philosophy adheres strictly to the principle of \textit{what you see is what you simulate}~\cite{muller2016simulating}. In contrast to the hybrid, two-part approach of VR-GS \cite{jiang2024vr}, where the visible Gaussian shells are driven by an invisible cage mesh, our framework can be considered as truly unified. The surface that the user sees and interacts with--defined by the GS representation--is directly coupled to the same underlying mesh that the physics engine simulates. This tight integration ensures that every interaction, no matter how subtle, has a direct and accurate visual and physical consequence. This unified approach not only enhances realism but also simplifies the content creation pipeline, as there is no need to generate or manage a separate physical proxy.

\section{Conclusion}

In this paper, we introduced \our{}, a novel mesh-based Gaussian Splatting (GS) system that enables physics-aware interaction with 3D content in VR. Unlike prior approaches that rely on simplified proxy geometries or tetrahedral cages \cite{jiang2024vr}, our method directly integrates surface meshes with Gaussian primitives, providing a unified, geometrically consistent representation. Such an approach supports both realistic rendering and physically plausible manipulation.

Through a comprehensive user study involving 18 participants and three distinct immersive scenes, we demonstrated that \our{} method delivers high usability, low cognitive workload, and strong engagement levels. Notably, participants rated the \textit{stretching} manipulation as significantly more natural than the state-of-the-art VR-GS baseline \cite{jiang2024vr}, while maintaining consistent performance across other physics-aware manipulation types, such as \textit{twisting} and \textit{shaking}. These results confirm the robustness of our approach and its ability to deliver stable, realistic, and intuitive interactions in complex VR environments.


\bibliographystyle{iclr2025_conference}

\end{document}